\UseRawInputEncoding
\pdfoutput=1
\documentclass[12pt]{iopart}

\usepackage{mathrsfs}
\usepackage{graphicx}
\usepackage{subfigure}
\usepackage{dcolumn}
\usepackage{bm}
\usepackage{amssymb}
\usepackage{amsmath}
\usepackage{hyperref}
\usepackage{paralist}
\usepackage{color}
\usepackage{pdfpages}
\usepackage{graphicx}
\usepackage{ulem}
\usepackage{iopams}

\newcommand{\normr}{\mathrm{normr}}
\begin{document}

\title{Continuous phase transition induced by non-Hermiticity in the quantum contact process model} 
\author{Wen-Bin He \footnote{These two authors contributed equally to the work}}
\address{Beijing Computational Science Research Center, Beijing, 100193, China}
\address{Quantum Systems Unit, Okinawa Institute of Science and Technology Graduate University, 904-0495 Okinawa, Japan}

\author{Jiasen Jin \footnote{These two authors contributed equally to the work}}
\address{School of Physics, Dalian University of Technology, Dalian 116024, China}

\author{Fernando Iemini}

\address{Instituto de F{\' i}sica, Universidade Federal Fluminense, 24210-346 Niter{\' o}i, Brazil}
\address{The Abdus Salam International Center for Theoretical Physics, Strada Costiera 11, 34151 Trieste, Italy.}

\author{Hai-Qing Lin}

\address{Beijing Computational Science Research Center, Beijing, 100193, China}
\address{School of Physics, Zhejiang University, Hangzhou, 310030, China.} 

 \eads{\mailto{fernandoiemini@gmail.com}, \mailto{haiqing0@csrc.ac.cn}}
\ead{submissions@iop.org}

\begin{abstract}
Non-Hermitian quantum system recently have attracted a lots of  attentions theoretically and experimentally. However, the results based on the single-particle picture may not apply to understand the property of non-Hermitian many-body system. How the property of quantum many-body system especially the phase transition will be affected by the non-Hermiticity remains unclear. Here we study non-Hermitian quantum contact process (QCP) model, whose effective Hamiltonian is derived from Lindbladian master equation. We show that there is a continuous phase transition induced by the non-Hermiticity in QCP. We also determine the critical exponents $\beta$ of order parameter, $\gamma$ of susceptibility and study the correlation and entanglement near phase transition point. We observe that the order parameter and susceptibility display infinitely singularity even for finite size system, since non-Hermiticity endow many-body system with different singular behaviour from classical phase transition. Moreover our results show that the phase transition have no counterpart in Hermitian case and belongs to completely different universality class. 
\end{abstract}

%
%
%
%
%

\maketitle

\section{Introduction} 
Non-Hermitian physics currently has attracted considerable attentions \cite{Ueda_review,Bergholtz}, which greatly extend the knowledge about quantum system. For example, the complex energy coalesce at exceptional point(EP)\cite{Bergholtz,Hatano,wang18,Robin}; The wave function localize at the edge of quantum system, which is now named as non-Hermitian skin effect\cite{Hatano,wang18,Gong20,PXue18}. Actually, quantum systems almost are subjected to environment. It is more appropriate to consider the quantum system to be  non-Hermitian(NH). Recently, it was found that driven-dissipative non-Hermitian system can provide one method to control quantum system, for instance, directional amplification  \cite{Clara_nc,Clara_l,Porras19,Porras21}. Non-Hermiticity can impose great influence on the properties of the topological non-trivial system. It needs generalized Brillouin zone to establish a generalized Bloch band theory \cite{wang18,ZhongWang,kazuki}. Non-Hermiticity can also induce phase transition without gap\cite{Norifumi} and non-Hermitian many-body localization \cite{Hamazaki19}. Non-Hermitian quantum system needs more effort to be studied, understood and utilized.
  
Though we have understood several interesting property of non-Hermitian quantum system which based on single-particle picture, it is an important issue to understand that how non-Hermiticity affect many-body system, especially the phase transition of many-body system. The theory of phase transition provide people profound understanding to the status of the matter. When non-Hermiticity comes into quantum many-body system, whether non-Hermiticity can induce phase transition, and if the phase transition exists  how to classify this type of phase transition. These questions remain being elusive. Moreover, what the relation is between the phase transition induced by non-Hermiticity and their Hermitian counterpart. As the primary motivation to answer above questions, we are going to explore the phase transition induced by non-Hermiticity in quantum contact process model.

In this work, we consider the open system whose dynamics governed by Lindbladian master equation. We use effective Hamiltonian to describe the property of non-Hermitian quantum system. Then we choose quantum contact process(QCP) model as research object, which breaks $U(1)$ symmetry. We combine analytical method for special case of spin number $L=2$ and exact diagonalization(ED) for general $L$ to compute energy spectrum, the order parameter $M^{x}$, susceptibility $\chi$, correlation, and entanglement entropy. The results of this work show that there is continuous phase transition within the non-Hermitian QCP model. We extract the critical exponents, $\beta$ of order parameter, $\gamma$ of susceptibility. We reveal the transition from quasi long-rang order to short-range order and entanglement properties near critical point. Finally, we compare the results of non-Hermitian QCP model with Hermitian counterpart, which shows the phase transition in non-Hermitian QCP model have no correspondence in Hermitian case. By looking at the critical exponents, the results in our work may indicate the phase transition belongs a new universality class. Moreover, even for finite size system, we notice non-Hermiticity endow many-body system with different singular behaviour from classical phase transition.  Our results can extend the knowledge about the phase transition of non-Hermitian quantum matter.

\section{Non-Hermitian QCP model}
For the open quantum many-body systems, its dynamics usually can be determined by Lindbladian master equation \cite{breuer} as
\begin{equation}
\dot{\rho}(t)=-i[\hat{H}_{0},\rho(t)]+\sum_{k}^{L}{{\cal D}_{k}[\rho(t)]}.
\label{ME}
\end{equation}
where $\hat{H}_{0}$ is Hamiltonian governing the coherent evolution of the system. The dissipation superoperator ${\cal D}_{\alpha}[\rho(t)]=
\hat{L}^{k}\rho(t)\hat{L}^{k,\dagger}-\frac{1}{2}\{\hat{L}^{k^\dagger} \hat{L}^{k},\rho(t)\}$ is defined by the following Lindbladian jump operator,  $\hat{L}^{k}=\sqrt{\Gamma} \sigma_{-}^{k}$ which impose local spin decaying channels. In the absence of quantum jumps during a time interval, the term $L^k\rho(t)L^{k\dagger}$, which describes the state transition, can be omitted\cite{Ueda_review,Andrew}. This corresponds to postselection \cite{Andrew,Brun}in the experiment, which target on succeed measurement results. In this case, the quantum system is considered to undergo a non-unitary time evolution governed by an effective Hamiltonian which is derived from the master equation(\ref{ME}) as follows
\begin{equation}
\hat H_{\rm eff} = \hat H_{0}-\frac{i}{2}\sum_{k=1}^{L} \hat{L}^{k^\dagger} \hat{L}^{k}. 
\end{equation}
In following discussion, we omit subscript for the non-Hermitian Hamiltonian. Since the Lindbladian operator above, we are interested in exploring the properties of non-Hermitian many-body system which is described by the effective Hamiltonian as
\begin{equation}
\hat H=\hat H_{0}-\frac{i}{2} \Gamma \sum_{k=1}^{L} \hat \sigma_{+}^{k}\hat \sigma_{-}^{k}.
\label{Heff_H0}
\end{equation}
From above formula, which gives out the non-Hermitian Hamiltonian of a class of spin many-body system with jump operator of $\hat{L}^{k}=\sqrt{\Gamma} \sigma_{-}^{k}$, it can be found that dissipation imposes imaginary field along $z$-direction by looking at formula $\sigma_{+}^k\sigma_{-}^k=(\sigma_{z}^k+I)/2$. We look back at the knowledge about the general matrix.
There are the left and right eigen-energy and eigenvectors with the non-Hermitian effective Hamiltonian, which are denoted by
\begin{eqnarray}
\hat H |\phi_{R}^i \rangle = E_i | \phi_{R}^i \rangle, \\
 \langle \phi_{L}^i | \hat H = E_i \langle \phi_{L}^i|,
\end{eqnarray}
with eigenvalues $E_i \in \mathbb{C}$.  The left and right eigenvectors satisfy  bi-orthonormal relation and completeness relation \cite{Heiss}
\begin{eqnarray}
\langle \phi_{L}^{i}\vert\phi_{R}^{j} \rangle= \delta_{ij} \langle \phi_{L}^{j}\vert\phi_{R}^{j} \rangle, \\
\sum_{j} \frac{ \vert\phi_{R}^{j} \rangle \langle \phi_{L}^{j}\vert}{\langle \phi_{L}^{j} \vert\phi_{R}^{j} \rangle}=I.
\end{eqnarray}
The properties of eigen-energy and eigenvectors of non-Hermitian are completely different from the Hermitian case, which endow non-Hermitian system with interesting new physics. When at least two eigen-energies  coalesce with each other, exceptional point(EP) emerges\cite{Hatano,wang18,breuer,Heiss,Lee,Tony}. In order to obtain EP with our model in Eq(\ref{Heff_H0}), original Hamiltonian $H_{0}$ must not conserve spin polarization. If $\left[ \hat H_{0},\sum_{k} \sigma_{z}^k \right]=0$ holds, the non-Hermitian term in Eq(\ref{Heff_H0}) can only give us trivial complex energy. Such that we choose quantum contact process(QCP) model in 1D as research object
\begin{equation}
\hat{H}_{0}=\Omega \sum_{k=1}^L \left( \hat  \sigma_{x}^{k}\hat \sigma_{n}^{k+1}+\hat \sigma_{n}^{k} \hat  \sigma_{x}^{k+1} \right),
\end{equation}
with $\hat \sigma_{n}^k=\hat \sigma_{+}^k \hat \sigma_{-}^k$ and $L$ the number of spins in the system. QCP model has been widely studied in open many-body system, such as quantum epidemic dynamics \cite{Lesanovsky,Wintermantel}, experimental realization by Rydberg atom \cite{Gutierrez}.  We arrive at non-Hermitian Hamiltonian as
\begin{equation}
\hat H = \Omega \sum_{k}^{L} \left(  \hat \sigma_{x}^{k}\hat \sigma_{n}^{k+1}+\hat \sigma_{n}^{k}\hat \sigma_{x}^{k+1} \right)-\frac{i}{2} \Gamma \hat\sum_{k=1}^{L} \hat \sigma_{+}^{k} \hat \sigma_{-}^{k},
\label{Heff}
\end{equation}
which determines non-Hermitian physics in our model. For convenience with following discussion, we set $\Omega=1$ to make the parameter of Hamiltonian dimensionless. In our work, we always use periodic boundary condition.

\begin{figure}
\begin{center}
\includegraphics[scale=0.425]{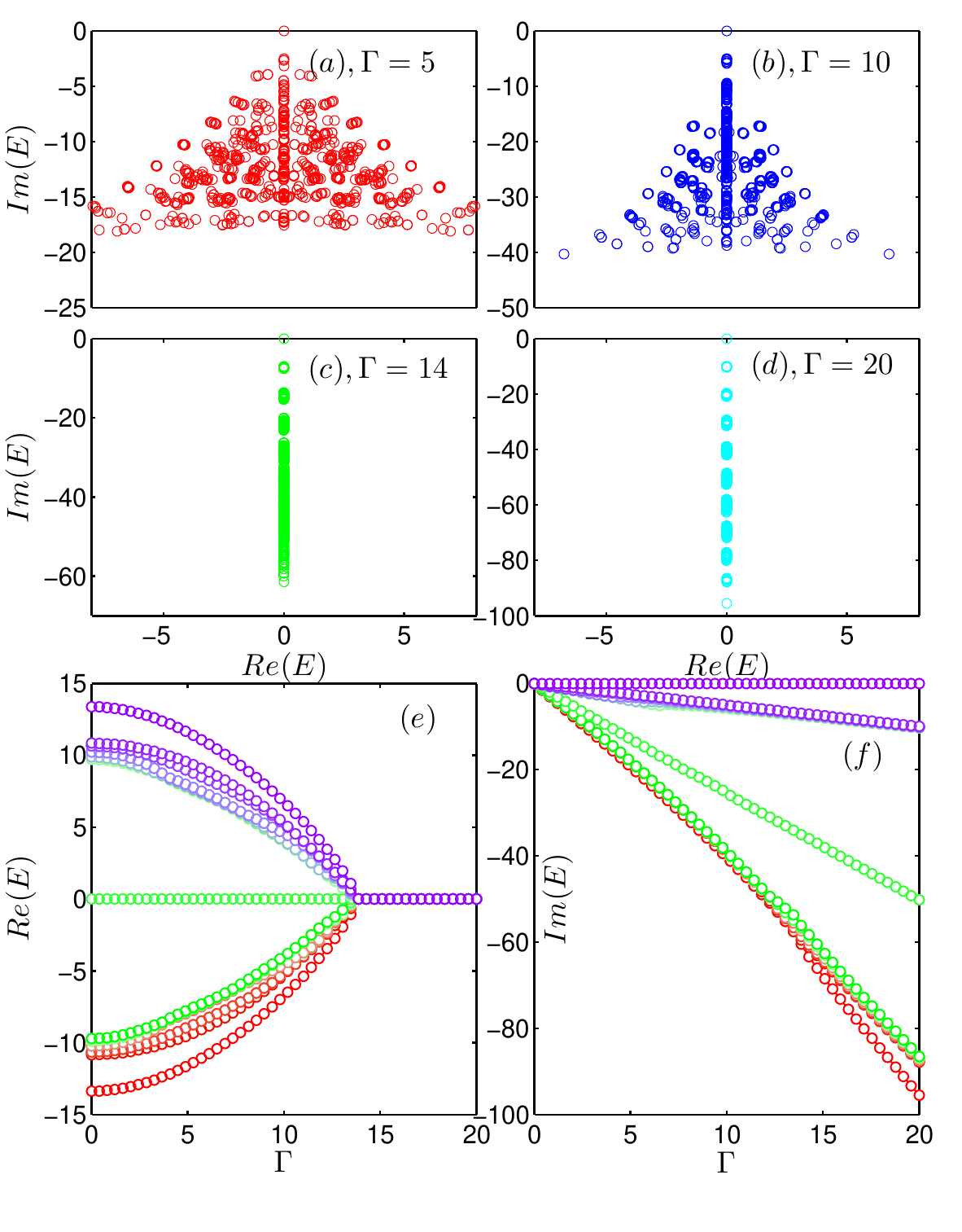}
\end{center}
\caption{The subplots (a)(b)(c)(d), distribution pattern of complex energy spectrum for four different $\Gamma$; The lowest and highest seven real part (e) and their imaginary part (f) of energy spectrum $E$ vary with $\Gamma$. $\Omega=1,L = 10$. }
\label{fig1_E_Gamma}
\end{figure}

We firstly look at the complex energy spectrum of the non-Hermitian system for two spins case $L=2$. We obtain four eigen-energies as $E_{1}=0 \quad (\rm trivial)$, $E_{2}=-i\Gamma/2$, $E_{3,4}=\frac{-i3\Gamma}{4} \pm \frac{1}{4} \sqrt{ 32\Omega^2-\Gamma^2}$, please see the \cite{sm} for the matrix and eigenvectors for $L=2$. It is easy to determine the EP as $\Gamma_{c}=\sqrt{32} \Omega$, where $E_{3}$ and $E_{4}$ coalesce with each other. For many-body case, we look at the energy spectrum changing with dissipation strength $\Gamma$ by exact diagonalization(ED). We used brute-force diagonalization up to $N=12$ as {\it eig of matlab}, and diagonalization based on Lanczos method for ground state up to $N=16$ as {\it eigs of matlab}. See Fig. \ref{fig1_E_Gamma} subplots(a-d), energy spectrum changes with $\Gamma$ from distributing on complex plane of $E$ to distributing on imaginary axis. Moreover, the real part of the energy varies with dissipation strength $\Gamma$ from finite values to zero at around $\Gamma \sim 14$, whose singularity looks like the order parameter of second-order phase transition \cite{Sachdev}; while there is split with the imaginary part of energy.  We also notice that $i \hat H_{eff}$ satisfies pseudo-Hermiticity \cite{Ueda_review} by transforming Hamiltonian and energy spectrum
\begin{align*}
&\hat H_{eff} \rightarrow i \hat H_{eff} \\
&(\pm E_{re}+i E_{im}) \rightarrow  (\pm i E_{re}- E_{im})
\end{align*}
which results in that the eigenvalues appear in complex conjugate pairs at left of EP and the spectrum are entirely real at right of EP. These analysis show that $H_{eff}$ of NH QCP has anti-PT symmetry breaking \cite{Ramy,ozdemir,YYang} near EP. The singularity of the real part energy and anti-PT breaking indicates a transition in the energy spectrum of the non-Hermitian model (\ref{Heff}), as a complex-imaginary transition.

In order to understand the complex-imaginary transition better, we are going to analyse the phase transition by studying the order parameter and susceptibility. But here we need to make preparation by defining the ground state of non-Hermitian many-body system.  {\it Ground  states: } we define the "ground state" $|\psi_{R/L}\rangle$ as the state $\psi_{0}$ associated to the energy $E_0$ with minimum real part as did in literatures \cite{Norifumi,weitao}. The state with smallest real part can map to traditional ground state through continuation between the non-Hermitian case and Hermitian case, which can be observed within short time evolution in the absence of quantum jump\cite{Hamazaki19,Norifumi} due to its imaginary part. The state with largest imaginary part corresponds to the steady state that can be observed within long time evolution \cite{Tony,Lee}. By continuation we mean that the non-Hermitian Hamiltonian becomes the Hermitian case when the dissipation $\Gamma$ tends to zero and the chosen "ground state" recover to the traditional one, namely the state corresponds to the smallest real energy. The former satisfies the requirement of continuation while the latter does not. When the system crosses to right side of the exceptional point, since energies become imaginary such that above definition of ground state become invalid, we track the continuity and analyticity of energy and observables to choose the ground state. Concretely, the ground state should keep the continuity of imaginary part of ground state from left side to right side of EP firstly. After reaching right side of EP, we use continuity and analyticity of observables to determine the ground state. In following part, we will reveal the phase transition induced by non-Hermiticity in our model in detail.

\begin{figure}
\begin{center}
\includegraphics[scale=0.45]{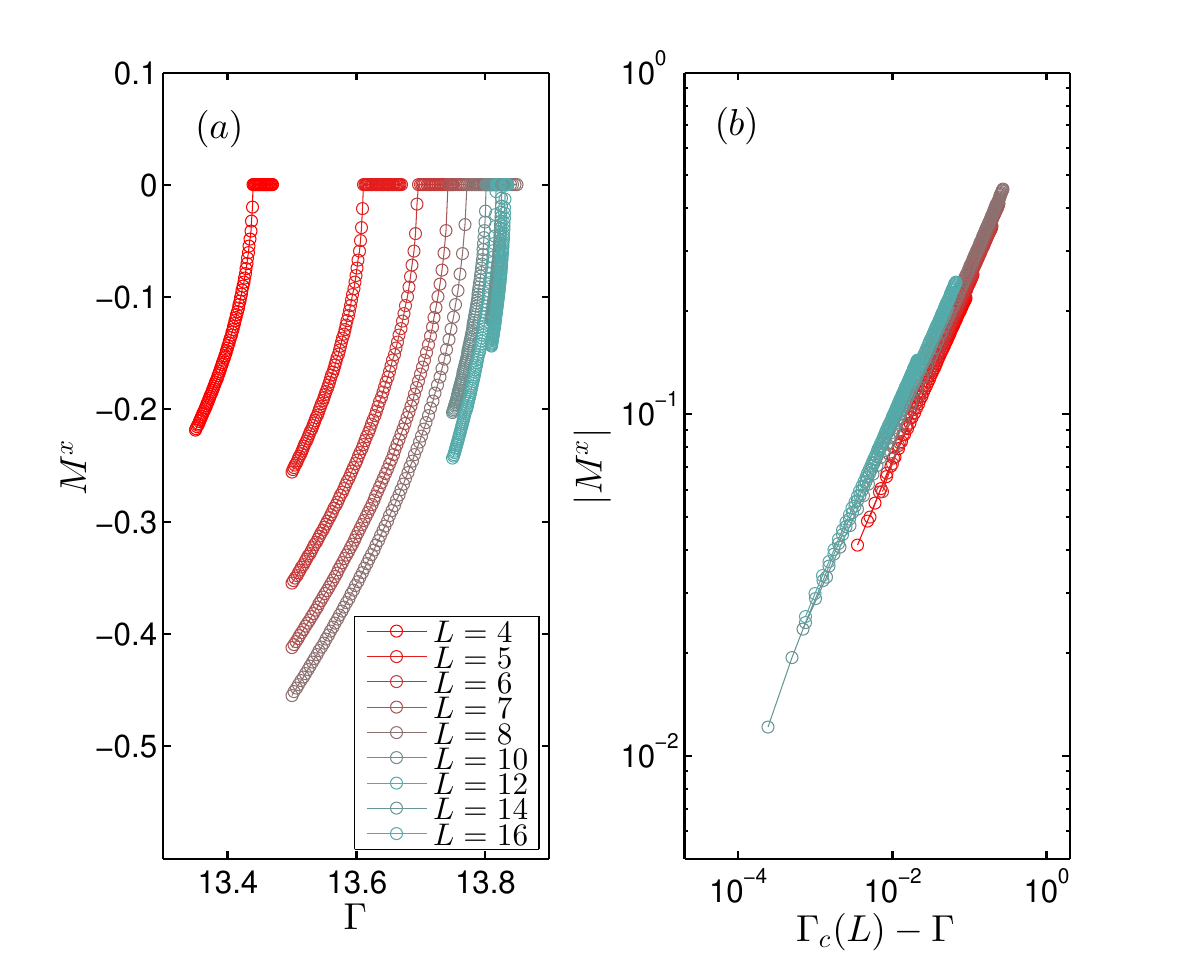}
\end{center}
\caption{(a)The spin polarization  along $x$-axis direction $M^{x}=\left\langle \sum_{k}\sigma^{k}_{x} \right\rangle$ of ground state vary with $\Gamma$  for different system size $L$; (b) the absolute value of $M^{x}$ near critical point V.S dissipation strength in logarithmic scale. The scaling exponent of $M^{x}\sim -|\Gamma_{c}(L)-\Gamma|^{\beta}$ is fitted out for different system size, see the Table.(\ref{tab_mx}). $ \Omega=1$. }
\label{fig_mx}
\end{figure}

\begin{table}
\centering
\caption{The critical points,critical exponents of order parameter of spin polarization along x-direction $M^{x}$ for different system size $L$. $\normr$ stands for norm of residuals.}
\begin{tabular}{|p{0.8cm}|p{1cm}|p{1cm}|p{1cm}|p{1cm}|p{1cm}|p{1cm}|p{1cm}|p{1cm}||cccccccc}
\hline
\hline
$L$& 4 & 5 &6 &7 &8 &9 &10    \\
\hline
$\Gamma{c}(L)$&13.4396& 13.6116& 13.6969& 13.7424& 13.7720& 13.7898& 13.8016  \\
\hline
$\beta$& 0.5102& 0.5209& 0.5303& 0.5117& 0.5261&0.5251&0.5026 \\
\hline
$\normr$ & 0.0336& 0.0582& 0.0953& 0.0108& 0.0319& 0.0383& 0.0122\\
\hline
$L$&11 &12 & 13 &14 &15 & 16&   \\
\hline
$\Gamma{c}(L)$& 13.8106& 13.8169 &13.8223& 13.8259 &13.8289 & 13.8312&   \\
\hline
$\beta$& 0.5069 & 0.4968 & 0.5178 & 0.5120 & 0.5201 & 0.5122 &    \\
\hline
$\normr$ & 0.0111 & 0.0687 & 0.0063 & 0.0129 & 0.0504  & 0.0181 &    \\
\hline
\hline
\end{tabular}
\label{tab_mx}
\end{table}

\section{The observables and critical exponents}
Generally, we use right ground eigenstate to compute the observables
\begin{equation}
O=\langle \psi_{R}\vert  \hat{O} \vert  \psi_{R}\rangle,
\end{equation}
which can make sure that observables are always real. In fact, it can give the same results by using right eigenstate and using left-right eigenstate, see Fig.\ref{fig_chiLR} in appendix \cite{sm}.
In order to compute the susceptibility of system, we impose a small probe magnetic field along the $z$-axis,
\begin{equation}
\hat H \rightarrow \hat H-\delta h/2 \sum_{k}\hat \sigma_{z}^{k}=\hat H-\delta h  \sum_{k}(\hat \sigma_{+}^{k} \hat \sigma_{-}^{k} -I/2).
\end{equation}
The spin polarization $M^{\alpha }=\langle \psi_{R}\vert \sum_{k} \hat \sigma_{\alpha }^{k}\vert  \psi_{R}\rangle$ for general direction $\alpha = x,y,z$. In our work, we choose spin polarization along $x$-axis $M^{x}$ as order parameter. The susceptibility can be given by the differential,
\begin{equation}
\chi= (M^{z}(\delta h)-M^{z}(0))/2 \delta h.
\end{equation}
The non-analytical behaviour of order parameter and susceptibility can help us to reveal the property of phase transition for non-Hermitian QCP model.

As shown in Fig. \ref{fig_mx}, order parameter $M^{x}$ vary with dissipation strength $\Gamma$ for different spin number up to $L=16$ in linear axis (a) and logarithmic axis (b). The critical points for different system size are determined when $M^{x}$ becomes zero. While in Fig. \ref{fig_mx}(b) the data almost collapse into one straight line in logarithmic scale, we extract the critical exponents $\beta$ for different spin number according to scaling function
\begin{equation}
M^{x}\sim -|\Gamma_{c}(L)-\Gamma|^{\beta}.
\end{equation}
The critical points for different system size gradually converge to true critical point. In Table.\ref{tab_mx}, we list critical points and critical exponents for system size from $L=4$ to $L=16$. And $\normr$ stands for norm of residuals, which quantifies the error for fitting our data to the scaling function in logarithmic axis. All critical exponents for different system sizes are nearly the same.  After taking average of critical exponents of different $L$, the critical exponent of order parameter is estimated as $ \bar{\beta}  \approx 0.51$.
In fact, we can look at theoretical  results of special case $L=2$ to check numerical results. The order parameter can be written as 
\begin{equation}
M^{x} \sim \sqrt{2 \Gamma_{c}(2)} \sqrt{\Gamma_{c}(2)-\Gamma} \sim \sqrt{\Gamma_{c}(2)-\Gamma},
\end{equation}
here $\Gamma_{c}(2)=\sqrt{32} \Omega$ mean the EP for $L=2$, see appendix \cite{sm} for details. This theoretical analysis confirm numerical results of critical exponent. From above analysis, we can confidently claim the true critical exponent for thermodynamic limit $\beta=1/2$ \cite{BoWei}. The results of order parameter $M^{x}$ confirm the analysis by energy spectrum in Fig. \ref{fig1_E_Gamma}, namely there is continuous phase transition induced by non-Hermiticity in QCP model. To be added, we observe the non-analytical behaviour of order parameter for finite size system even $L=2$. While, for classical phase transition of Hermitian models, non-analytical behaviour of order parameter appears in thermodynamic limit. When system size is finite, order parameter become smooth near critical point. In reference \cite{Lee}, the authors noticed similar behaviours.  The non-analytical behaviour of order parameter for finite system size in our work  may be related to non-Hermiticity, which is very different from classical phase transition of Hermitian case.  

\begin{table}
\centering
\caption{The critical points, critical exponents and amplitude of susceptibility for different system size $L$.}
\begin{tabular}{|p{1.2cm}|p{1.2cm}|p{1.2cm}|p{1.2cm}|p{1.2cm}|p{1.2cm}|ccccc}
\hline
\hline
$L$& 4 & 5 &6 &7 &8  \\
\hline
$\Gamma{c}(L)$&13.4388 &13.6092 &13.6949 &13.7418 &13.7704\\
\hline
$\gamma$& 1.5198 &1.5829 &1.4975 &1.5049 &1.4986  \\
\hline
$\chi_{0} *10^{-4}$ & 0.2972 &0.2614 &0.3377 &0.3384 &0.3573 \\
\hline
$\normr$ & 0.0009& 0.0061& 0.0001& 0.0001& 0.0003\\
\hline
$L$ &9 &10 &11 &12& \\
\hline
$\Gamma{c}(L)$ &13.7887 &13.8010 &13.8105 &13.8168& \\
\hline
$\gamma$&1.5273 &1.4969 &1.4999 &1.5731& \\
\hline
$\chi_{0} * 10^{-4}$ &0.3337 &0.3907 &0.3809 &0.2833& \\
\hline
$\normr$ & 0.0008& 0.0022& 0.0002& 0.0110& \\
\hline
\hline
\end{tabular}
\label{tab_chi}
\end{table}

\begin{figure}
\begin{center}
\includegraphics[scale=0.45]{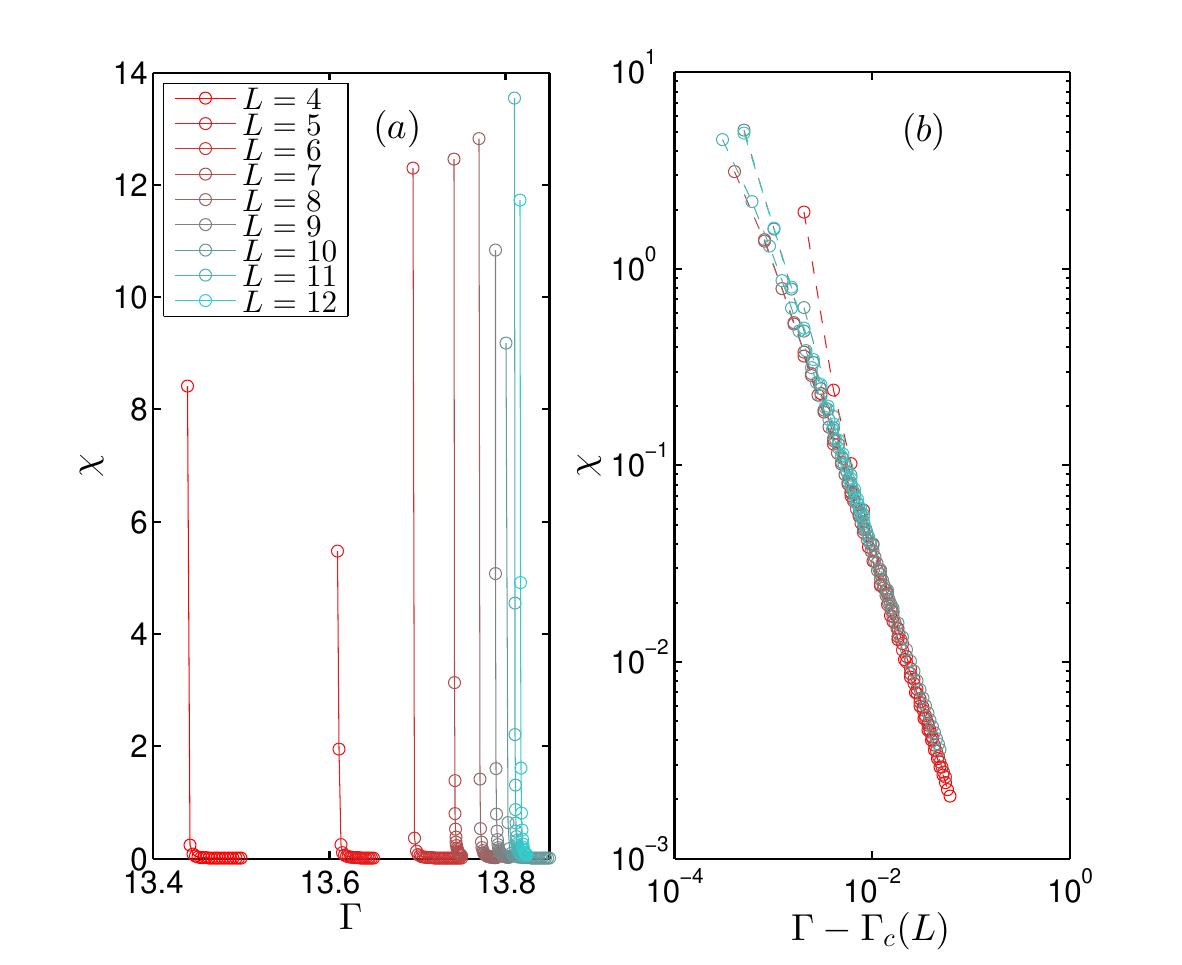}
\end{center}
\caption{ The susceptibility $\chi$ vary with  $\Gamma$ in linear scale(a) and vary with $\Gamma-\Gamma_{c}(L)$ in logarithmic scale(b) for different size $L=[4,5,6,7,8,9,10,11,12]$ . By linear fitting in logarithmic scaling of formula $\chi = \chi_{0}/(\Gamma-\Gamma_{c}(L))^{\gamma}$, the critical points, the scaling exponents $\gamma$ and amplitude $\chi_{0}$ in the table.\ref{tab_chi}.}
\label{fig_chi}
\end{figure}

We also study the magnetic susceptibility $\chi$ along $z$-direction, see Fig. \ref{fig_chi}. We show susceptibility vary with dissipation $\Gamma$ for different spin number up to $L=12$ in linear axis (a) and logarithmic axis (b). We can also observe that the susceptibility tends to infinite divergence for finite size system rather than finite peak for classical phase transition. As increasing system size, susceptibility strongly diverges near critical points in Fig. \ref{fig_chi}(a). Moreover  these diverging data can collapse into one straight line in logarithmic scale in Fig. \ref{fig_chi}(b) except the data of small size $L=4$ with small error. The critical exponent of susceptibility is extracted according to scaling relation
\begin{equation}
\chi = \chi_{0}/(\Gamma-\Gamma_{c}(L))^{\gamma}.
\end{equation}
In Table.\ref{tab_chi}, we list critical points, critical exponents and amplitude of susceptibility for different system size. By comparing the critical points of Table.\ref{tab_mx} and Table.\ref{tab_chi}, both results of $\Gamma_{c}(L))$ precisely agree with each other with negligible error. The critical exponents $\gamma$ are close to each other near $1.5$. After taking average of critical exponents $\gamma$, mean value of critical exponent is estimated as $\bar{\gamma}\approx 1.52$. Usually, EP is sensitive to disorder \cite{Yuce}. 
In appendix.C, we consider the non-Hermitian QCP model with random potentials and two spins case. The results indicate that the critical exponent changes even if under weak perturbation, the deviation from $\beta=0.5$ is slight within $5 \%$ as shown in Fig. \ref{nc2_qcp_nh_h0to1_beta} of Appendix.C

\begin{figure}
\begin{center}
\includegraphics[scale=0.38]{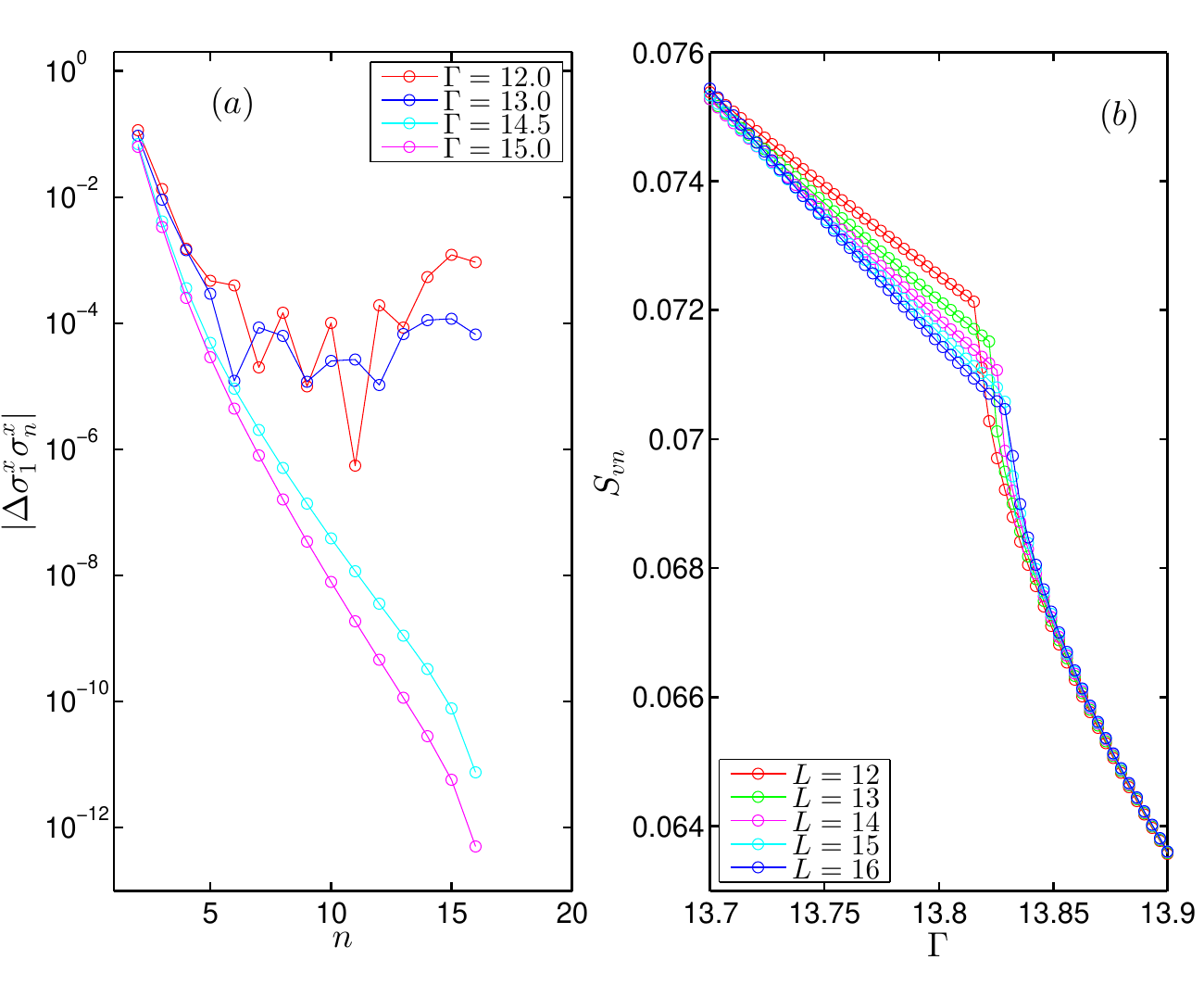}
\end{center}
\caption{ The correlation function $\Delta \sigma^{x}_{1}\sigma^x_{n}$ (a) vary with spin site $n$ for four different $\Gamma$ with system size $L=16$. Von Neumann entropy $S_{vn}$(b) vary with $\Gamma$ for half-partition of different system size $L$.}
\label{fig_corrS}
\end{figure}

\begin{figure}
\begin{center}
\includegraphics[scale=0.4]{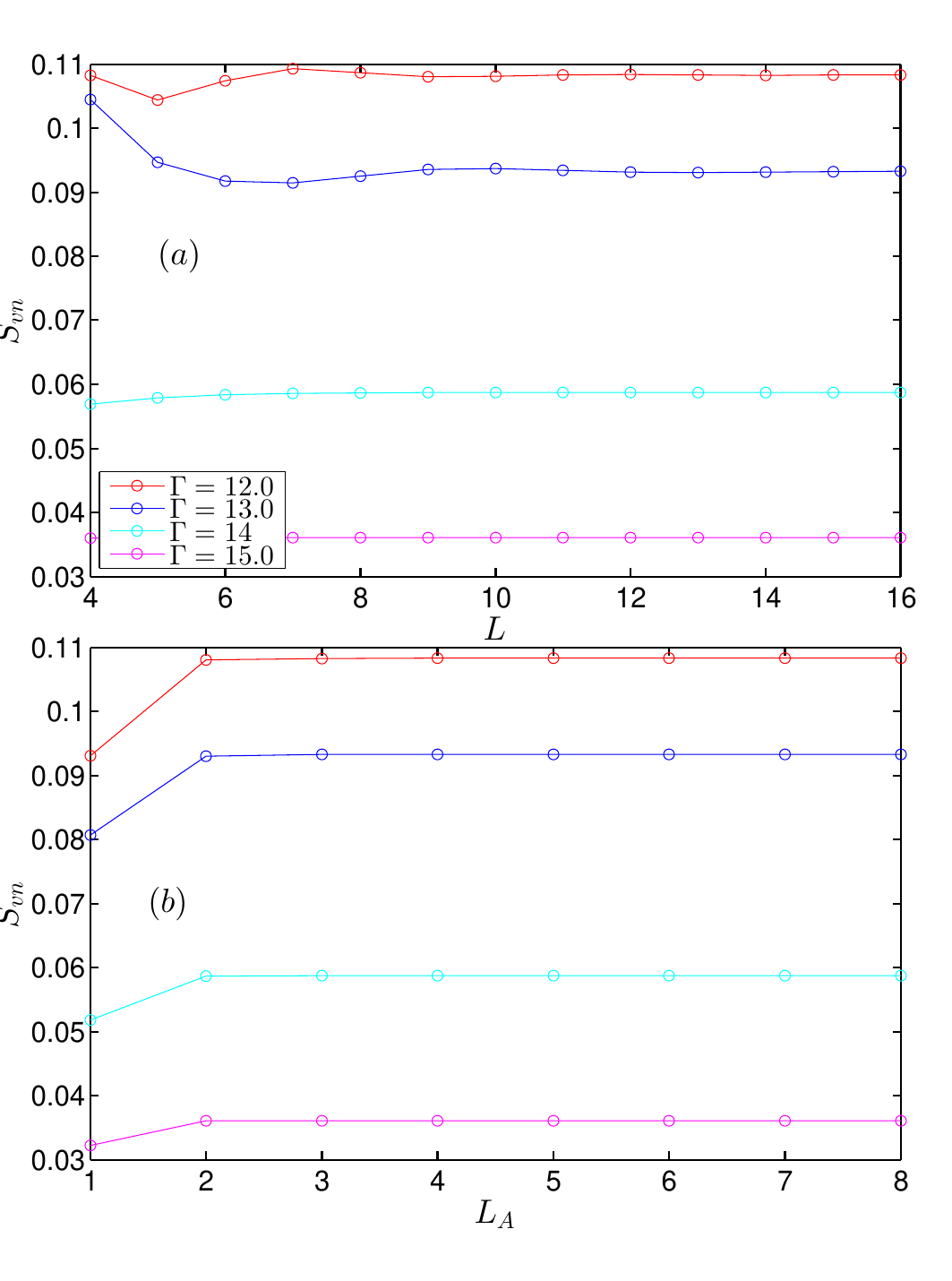}
\end{center}
\caption{Entanglement entropy $S_{vn}$ (a) for half-partition vary with system size $L$ and (b) for $L_{A}$-partition vary with $L_{A}$  with $L=16$. The dissipation strength $\Gamma=12.0,13.0,14,15.0$.  }
\label{fig_svn_LLA}
\end{figure}

\section{ Correlation and entanglement} 
In order to understanding the correlation and entanglement near phase transition, we study the correlation function of spin chain with open boundary
\begin{equation}
\Delta \sigma^{x}_{1}\sigma^x_{n}=\langle \hat \sigma_{x}^{1} \hat \sigma_x^{n} \rangle-\langle \hat \sigma_{x}^{1}\rangle \langle \hat \sigma_x^{n} \rangle.
\end{equation}
we also look at Von Neumann entropy $S_{vn}$ of subsystem-$A$ by partitioning system as $L=L_{A}+L_{B}$  
\begin{equation}
S_{vn}=-\Tr[\rho_{A} \ln \rho_{A}]= -\sum_{n}\lambda_{n}\ln \lambda_{n},
\end{equation}
where reduced density matrix of $A$ is $\rho_{A}=\Tr_{B}[\vert \psi_{R} \rangle \langle \psi_{R}  \vert]$ of ground state $\vert \psi_{R} \rangle$, and $\{ \lambda_{n} \}$ are the eigenvalues of $\rho_{A}$. During calculation, we define the half-partition as
\begin{equation}
L_{A}= \left\lbrace 
\begin{array}{ccc}
L/2 &, & \mod(L,2)=0\\ 
(L+1)/2& ,& \mod(L,2)=1\\
\end{array}
\right.
\end{equation}
we also used the right eigenstate to make sure that correlation function and Von Neumann entropy are real numbers in the non-Hermitian model.

As seen in fig.\ref{fig_corrS}(a), the correlation functions  vary with  spin site $n$ for parameters $\Gamma=12,13$(red and blue lines) at left side and $\Gamma=14.5,15$(cyan and pink lines) right side of critical point. Correlations have quasi long-range decay at left side of critical point and nearly exponential decay at right side of critical point. The system at left side seem be quasi long-range ordering (see similar case \cite{Lee}). Meanwhile, system at right side has short-range ordering.  The correlation at left and right side of critical point decay in completely different form regardless of boundary condition, as fig.\ref{fig_corr_pe&op} in appendix left panel for periodic boundary and  right for open boundary .

As seen in Fig.\ref{fig_corrS}(b), entanglement entropy decays with $\Gamma$ for different $L$. Non-analytical peaks indicate a NH phase transition for the system from a higher to a lower entangled state. For ground state of NH-QCP model, entanglement characterize the continuous phase transition by non-analytical peak. At same time, we consider two cases to study entanglement entropy, first one: $L$ changes from $4$ to $16$ for half-partition $L_A=L/2$ and second one: $L_A$ changes from $1$ to $L/2$ for fixed $L=16$. In Fig.\ref{fig_svn_LLA}, we also found the results that entanglement entropy $S_{vn}$ almost (a) do not change with $L$ for half- partition and (b) do not vary with $L_A$ for $L_A$ ranging from $1$ to $L/2$ when $\Gamma \in [12.0,13.0,14,15.0]$ is near $\Gamma_c$. These results reveal that entanglement entropy in NH-QCP model satisfies area law since the surface of spin chain is two spins as Hermitian model in 1D.

By summarizing above results, we have obtained the critical exponents $\beta$, $\gamma$ for the phase transition in non-Hermitian QCP model. As far as we know, $\beta$ in our model is same as mean field results; While $\gamma$ in our work are different from the known critical exponents. Since the phase transition is related with the coalescence of energy spectrum at EP, non-Hermiticity can endow non-Hermitian many-body system with new critical behaviour, such as shown in Fig. \ref{fig_mx}. \ref{fig_chi}. The analysis about critical exponents of $\beta$, $\gamma$ reveal that the phase transition in non-Hermitian QCP is independent of system size. But the critical points serve as function of system size $L$, and correlation length must be constrained by $L$. The critical behaviour is governed by the long-wavelength fluctuations such that the critical exponents are universal. For finite size, singularity of phase transition in this work accompany with finite correlation length. This peculiar behaviour should be derived from the singular behaviour of EP for non-Hermitian system. The phase transition for non-Hermitian many-body system may be beyond the classification for Hermitian many-body system.

\begin{figure}
\begin{center}
\includegraphics[scale=0.35]{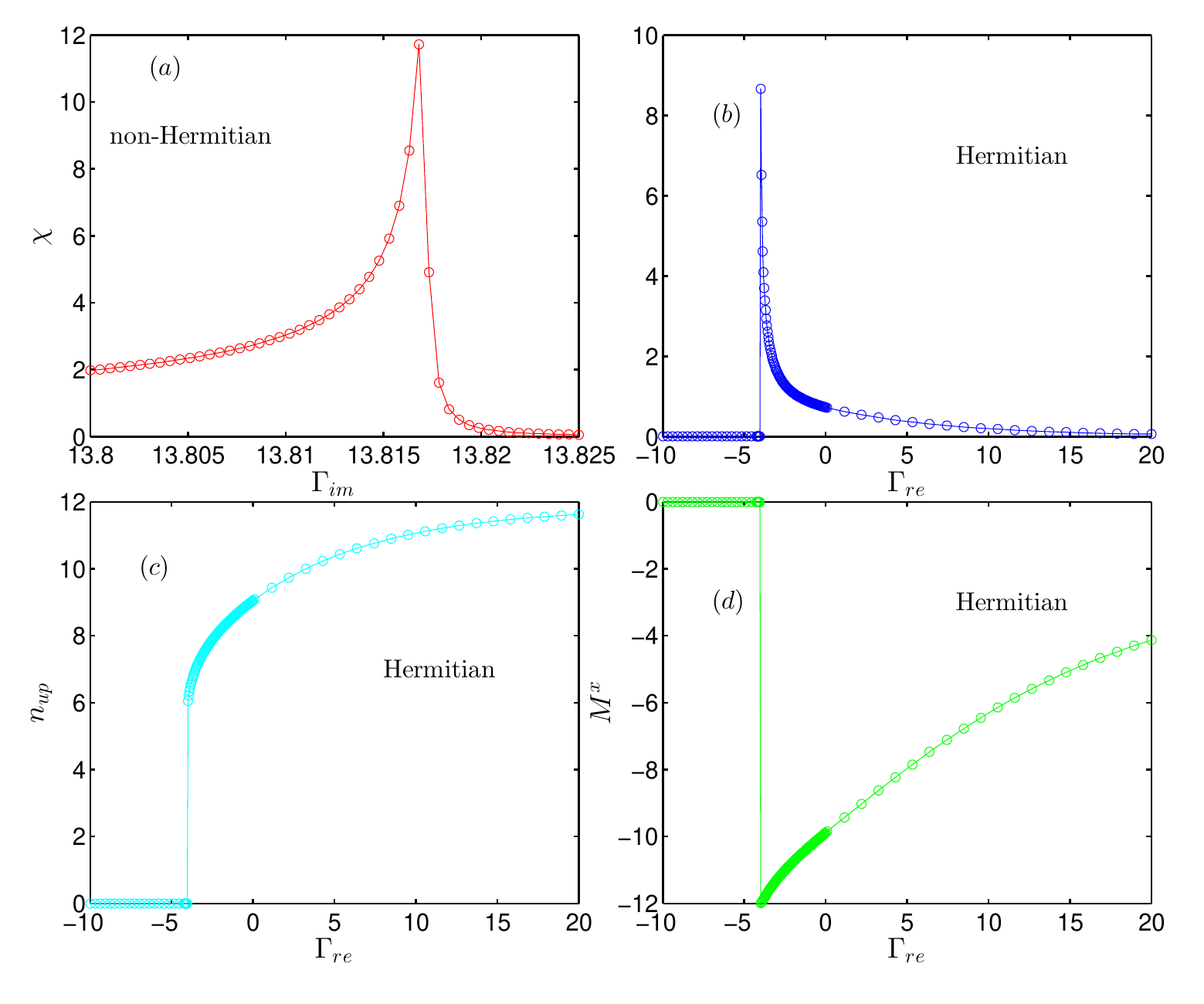}
\end{center}
\caption{subplot(a): By setting real part of $\Gamma$ as zeros, the susceptibility vary with $\Gamma_{im}$ for ground eigenstates of non-Hermitian QCP model. Subplot(b):By setting imaginary part of $\Gamma$ as zeros, the susceptibility of Hermitian QCP model vary with $\Gamma_{re}$ for ground eigenstates. The critical exponent for Hermitian model is $\gamma \sim 0.57$. Subplots(c)(d). The $n_{up}$ number of up-spin and order parameter $M^{x}$ vary with $\Gamma_{re}$. $L=12$, $ \Omega=1$.}
\label{figH&nH}
\end{figure}

\section{Comparison with Hermitian counterpart.} 
We compare the results of our model with its Hermitian counterpart by generalizing parameter to complex value $i\Gamma \rightarrow \Gamma_{\rm re}+i\Gamma_{\rm im}$ in Eq.(\ref{Heff}). The Hamiltonian becomes
\begin{equation}
\hat H=\Omega \sum_{k} \left(  \hat \sigma_{x}^{k}\hat \sigma_{n}^{k+1}+\hat \sigma_{n}^{k}\hat \sigma_{x}^{k+1} \right)-\frac{1}{2} (\Gamma_{\rm re}+i\Gamma_{\rm im}) \sum_{k=1}^{L}  \hat \sigma_{+}^{k}\hat \sigma_{-}^{k},
\label{qcpz}
\end{equation}
where we define $\Gamma=\Gamma_{\rm re}+i\Gamma_{\rm im}$. When $\Gamma$ taking real part $\Gamma_{\rm re}$, it corresponds to Hermitian case; While $\Gamma$ taking imaginary part $i\Gamma_{\rm im}$, it corresponds to non-Hermitian case in this work. In Fig.  \ref{figH&nH}, we show the susceptibility of non-Hermitian case (a) and Hermitian counterpart(b). By numerical fitting, the critical exponents $\gamma$ for both case are respectively near $1.5,0.57$. From the difference of critical exponent $\gamma$, it indicates that the continuous phase transition in non-Hermitian QCP model is different from its Hermitian counterpart. By looking at the number of up-spin(c)and order parameter(d) for Hermitian case, these observables discontinuously jump at critical point $\Gamma_{\rm re}^{c}\sim -4$, which imply first-order phase transition with Hermitian QCP model.

\begin{figure}
\begin{center}
\includegraphics[scale=0.4]{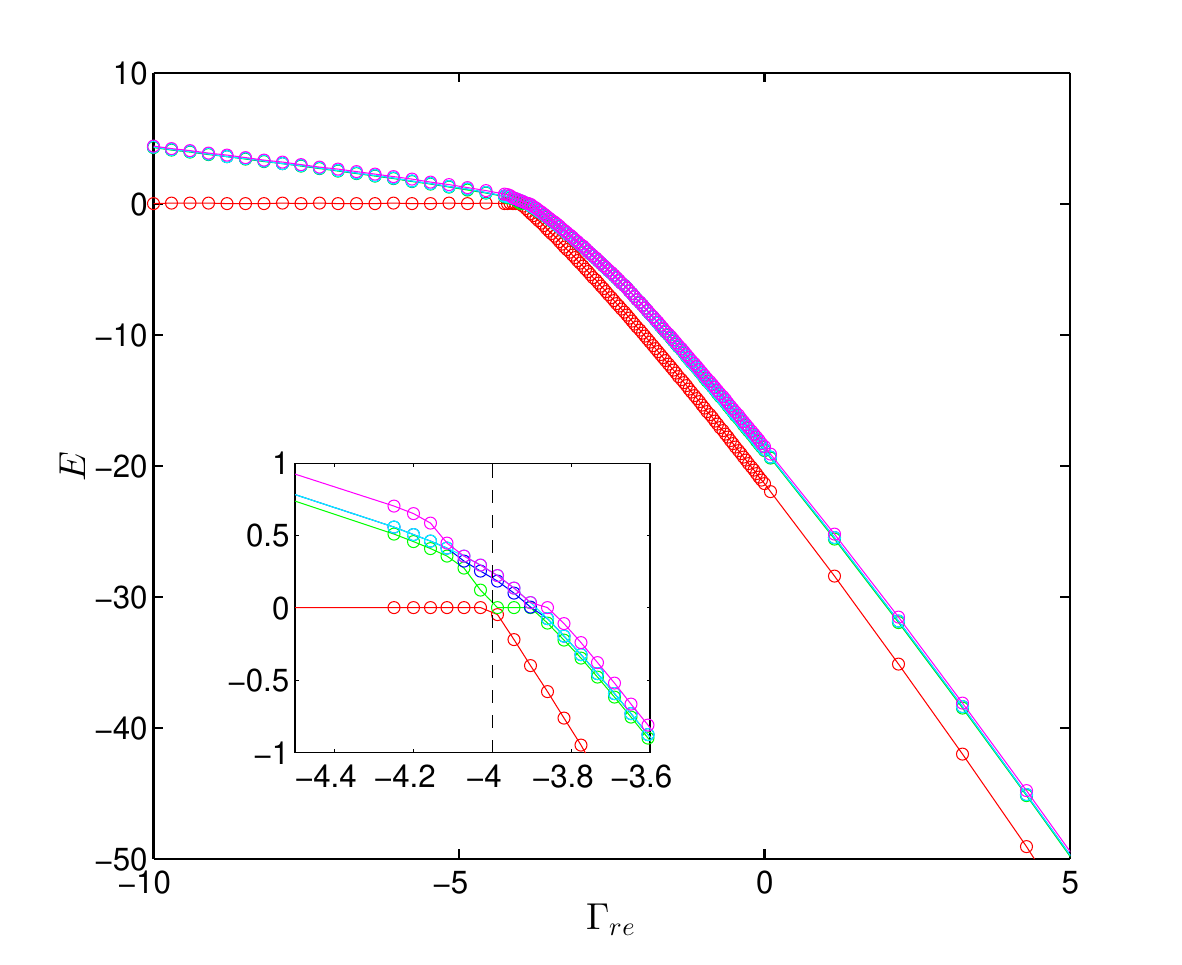}
\end{center}
\caption{The lowest five energy levels vary with real part of parameter $\Gamma_{re}$ for Hermitian case. $L=16$, $ \Omega=1$.}
\label{fig_E_hermL16}
\end{figure}

\begin{figure}
\begin{center}
\includegraphics[scale=0.37]{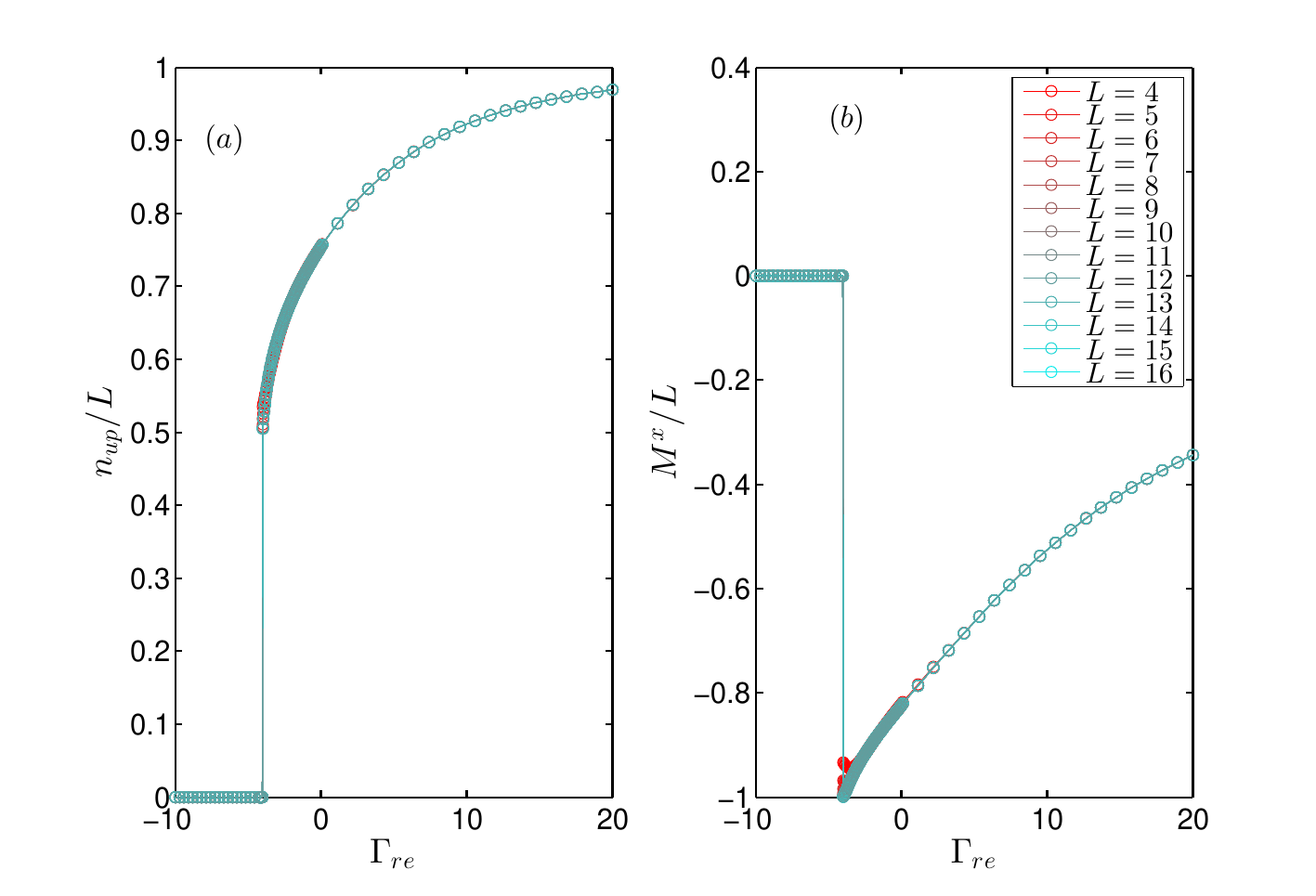}
\end{center}
\caption{The results for Hermitian model. subfig(a):The rescaled $n_{up}$ vary with $\Gamma_{re}$ for ground eigenstates of Hermitian QCP model for different system size $L$. subfig(b): rescaled $M^{x}$ vary with $\Gamma_{re}$ for different system size $L$. $ \Omega=1$.}
\label{fig_nupMx_hermL}
\end{figure}

We include complementary results of Hermitian model in Figs.\ref{fig_E_hermL16}.\ref{fig_nupMx_hermL}. As Fig.\ref{fig_E_hermL16} of numerical results shows, the lowest five energy levels vary with $\Gamma_{\rm re}$. There is level crossing point at $\Gamma_{\rm re}=-4\Omega$ for Hermitian QCP.  
As shown in Fig.\ref{fig_nupMx_hermL}, the number of up-spin and order parameter which are rescaled with $L$ vary with $\Gamma_{\rm re}$ for different system sizes,  indicating that the jump heights are proportional to system size $L$. There are sudden jumps with both observables near critical point. All results for different system size collapse into same curve. 

{\it Symmetry analysis:}
We transform the Hamiltonian of Hermitian case by using  $\sigma_{n}^k=(\sigma_{z}^k+1)/2$ 
\begin{eqnarray}
& \hat H &=\frac{\Omega}{2} \sum_{k} \left(  \hat \sigma_{x}^{k}\hat \sigma_{z}^{k+1}+\hat \sigma_{z}^{k}\hat \sigma_{x}^{k+1} \right)+\Omega \sum_{k}\hat \sigma_{x}^{k}  \nonumber \\
&-& \frac{1}{4} \Gamma_{\rm re} \sum_{k=1}^{L}  \hat \sigma_{z}^{k}+const.
\end{eqnarray}
We notice that in the interaction terms the $x$ and $z$ directions play a symmetric role. On the other hand, the external fields (second and third terms) break this symmetry.
The symmetric point is $\Gamma_{\rm re}=-4\Omega$, which agrees with critical point in Figs.\ref{fig_E_hermL16}.\ref{fig_nupMx_hermL} by numerics. We can conclude that order parameter $M^{x}$ of ground state has sudden jump at two side of critical point hence first-order phase transition with Hermitian counterpart.

In fact, by summarizing the results of critical exponent, we think that phase transition in non-Hermitian QCP model belongs to new universality class from previous known's. Non-Hermitian many-body system break the Hermitian conjugation symmetry. This type of phase transition is induced by the non-Hermiticity. Moreover, exceptional point is continuous and non-analytical behaviour mathematically for finite size system even $L=2$. At EP, non-Hermiticity alters the property of ground state hence the non-analytical behaviour of order parameter and singularity of susceptibility. It is uniqueness of non-Hermitian system comparing with regular phase transition of Hermitian counterpart.

\section{Conclusion}
We have studied the phase transition in non-Hermitian QCP model. We have determined the critical exponents $\beta$, $\gamma$ which may indicate that the phase transition in non-Hermitian QCP model belongs to new universality class. Our results can help people to understand the property of non-Hermitian many-body system. Moreover, our results can stimulate more study about the phase transition of  non-Hermitian many-body system. As for outlook, the dynamics of non-Hermitian system will be interesting topic since non-unitary evolution is completely different from the unitary evolution of Hermitian model, like \cite{Haga,Orito}. Renormalization group can be used to  reveal the difference and relation of phase transition between the Hermitian model and non-Hermitian model \cite{Kawabata1}. It may be feasible to utilize Rydberg atom\cite{Gutierrez,Browaeys}  to simulate the non-Hermitian QCP model in our work to observe the continuous phase transition experimentally even for simple case of two spins. 

\section{Acknowledgements.}
 We gratefully thank Profs.Rosario Fazio and Stefano Chesi for their inspiring discussions and suggestions. W.B.H. acknowledges support from NSAF U1930402. J.J. acknowledges support from the National Natural Science Foundation of China (NSFC) via Grant No. 11975064. F.I. acknowledges the financial support of the Brazilian funding agencies National Council for Scientific and Technological Development-CNPq(Grant No. 308205/2019 − 7) and FAPERJ (Grant No. E-26/211.318/2019 and E-26/201.365/2022). H.Q. Lin acknowledges financial support from NSAF U1930402 and NSFC 12088101 and computational resources from the Beijing Computational Science Research Center. 
\section{Data availability}
The data that support the findings of this study are available upon reasonable request from the authors.

\appendix

\section{The notations and observables of non-Hermitian QCP}
In this part, we introduce the notations used in main text and show more supplementary results.

\begin{figure}
\begin{center}
\includegraphics[scale=0.4]{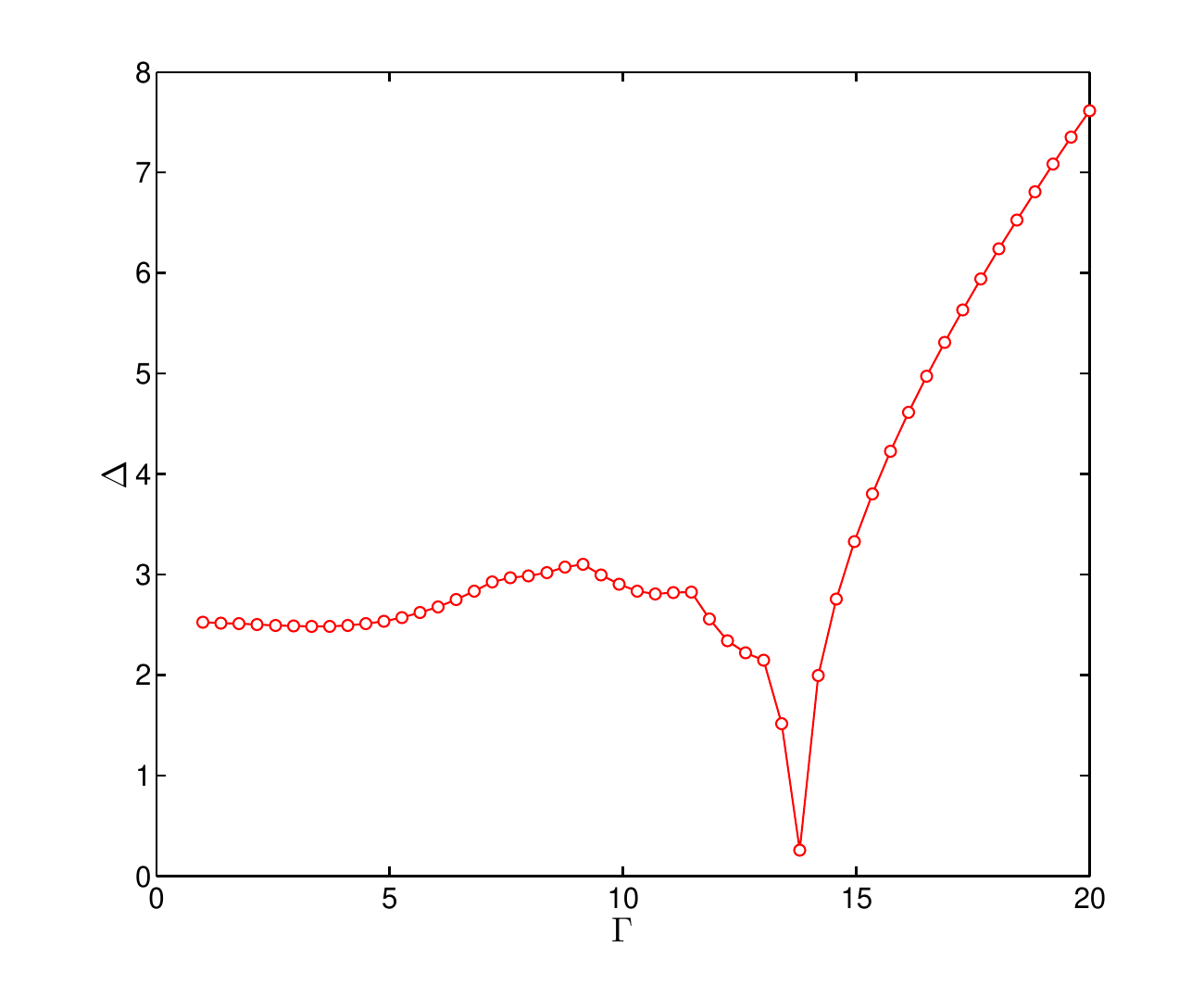}
\end{center}
\caption{The energy gap $\Delta$ vary with $\Gamma$ for ground state $L=10$, $ \Omega=1$.}
\label{fig_Egap}
\end{figure}

{\it Energy gap. } We define the energy gap as  following,
\begin{equation}
\Delta=\min_{n \neq 0} \vert E_{n}-E_{0} \vert.
\end{equation}
In Fig.\ref{fig_Egap}, the energy gap for no-Hermitian Hamiltonian vary with dissipation strength for $L=10$, which indicates the gap closing at critical point.

{\it Observables.}
 Generally, we use right eigenstate of ground state to compute the observables:
\begin{equation}
O=\langle \psi_{R}\vert  \hat{O} \vert  \psi_{R}\rangle.
\label{O_R}
\end{equation}
such as, number of spin-up
\begin{equation}
n_{\rm up} =\langle \psi_{R}\vert \sum_{k}  \hat \sigma_{+}^{k} \hat \sigma_{-}^{k}\vert  \psi_{R}\rangle.
\end{equation}
In this way, the observables and susceptibility are always real which is conformal to experimental detection. We can also use the left and right eigenstates of ground state to define the observable
\begin{equation}
O_{LR} =\langle \psi_{L}\vert  \hat{O} \vert  \psi_{R}\rangle.
\label{O_RL}
\end{equation}
Such as, magnetization $M^{\alpha }=\langle \psi_{R/L}\vert \sum_{k}
\hat \sigma_{\alpha }^{k}/2\vert  \psi_{R}\rangle$ for $\alpha = x,y,z$.

\begin{figure}
\begin{center}
\includegraphics[scale=0.35]{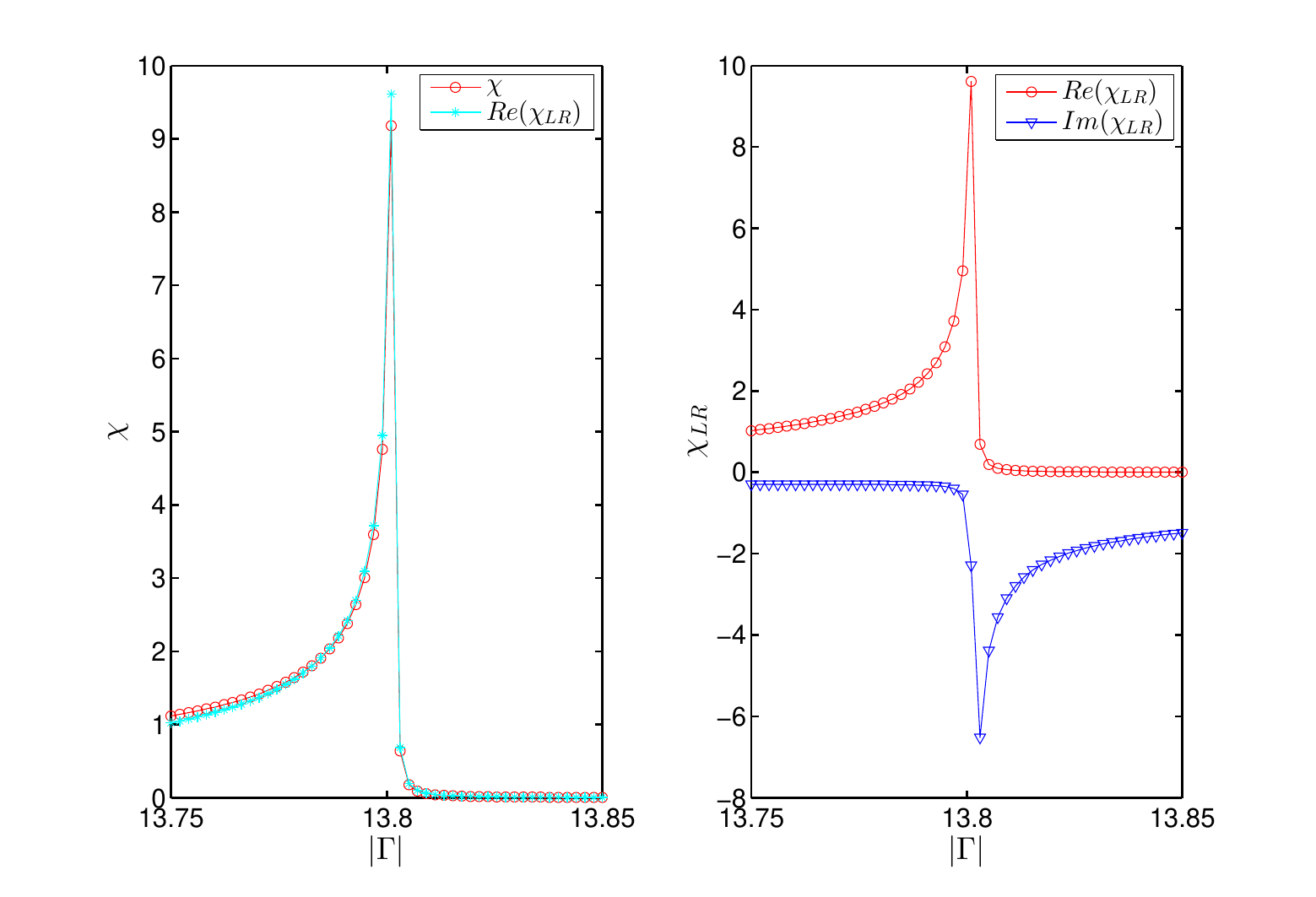}
\end{center}
\caption{subfig(a):The susceptibility vary with $\Gamma$ for right ground eigenstates of non-Hermitian QCP model.  subfig(b):The susceptibility vary with $\Gamma$ for right and left ground eigenstates of non-Hermitian QCP model. $L=10$, $ \Omega=1$.}
\label{fig_chiLR}
\end{figure}

We denote the susceptibility computed by two formulas (\ref{O_R}) and (\ref{O_RL}) as $\chi$ and $\chi_{LR}$ respectively. We think that $\chi$ and $\chi_{LR}$ both have the same function to characterize the property of phase transitions, see Fig.\ref{fig_chiLR}. Near critical point, $\chi$ is very close to the real part of  $\chi_{LR}$. 

\begin{figure}
\begin{center}
\includegraphics[scale=0.35]{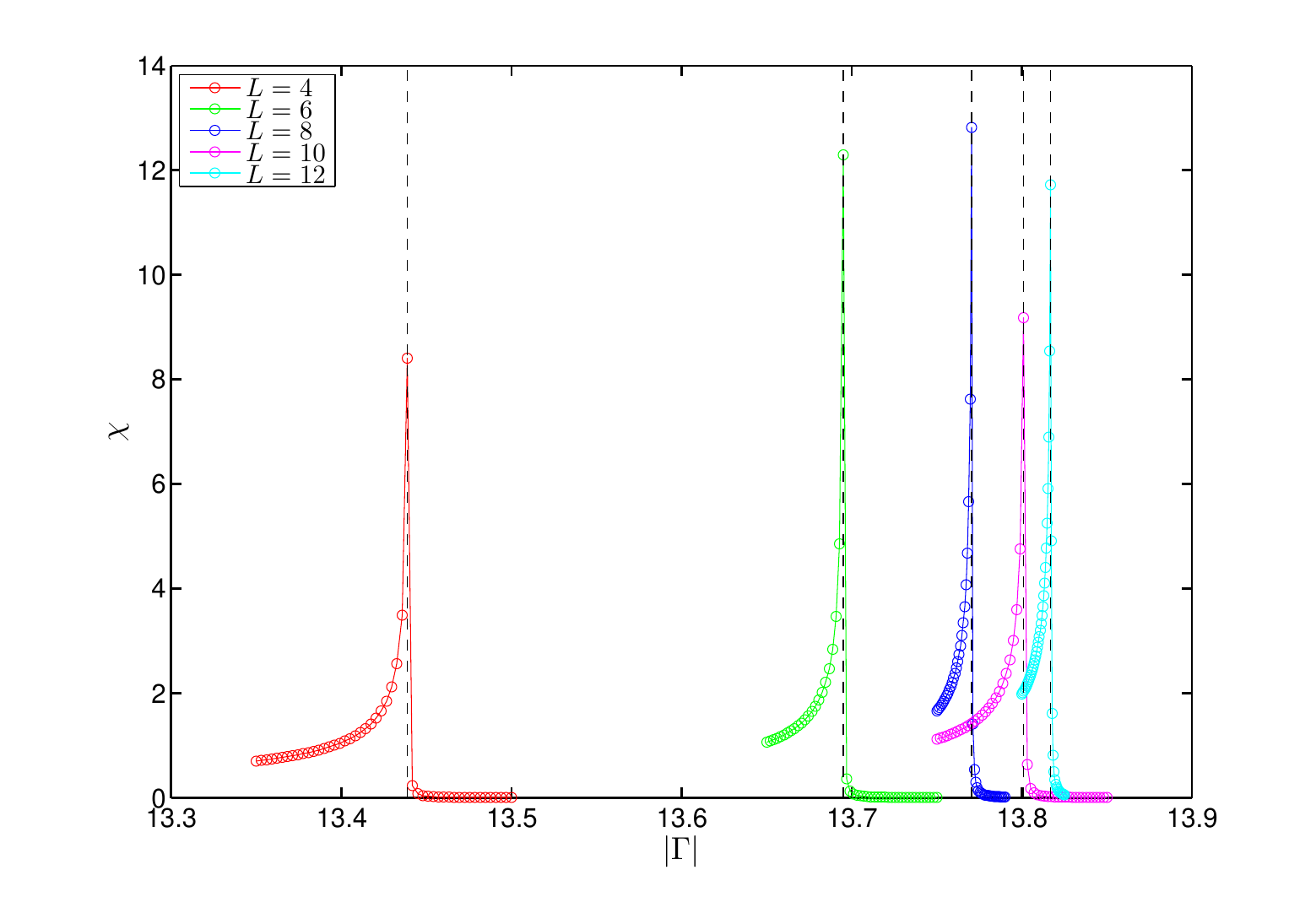}
\end{center}
\caption{ The susceptibility  vary with $\Gamma$ for ground eigenstates  for different $L$, $ \Omega=1$. The black dash lines indicate the critical points $\Gamma_{c}(L)$ of each spin number,  which are $13.4388,13.6949,13.770,13.801,13.8168$. We assume susceptibility satisfy scaling law $\chi \sim 1/(\Gamma-\Gamma_{c})^{\gamma}$. By using the right data of $\Gamma_{c}(L)$ which are larger than the critical point of each cluster size, the critical exponents of each cluster can be extracted. }
\label{fig_chiL}
\end{figure}

In Fig.\ref{fig_chiL}, we show susceptibility $\chi$ vary with dissipation strength for different system size. The singular point help us determine the critical points as shown in Tabs. \ref{tab_mx}.\ref{tab_chi}.

\begin{figure}
\begin{center}
\includegraphics[scale=0.35]{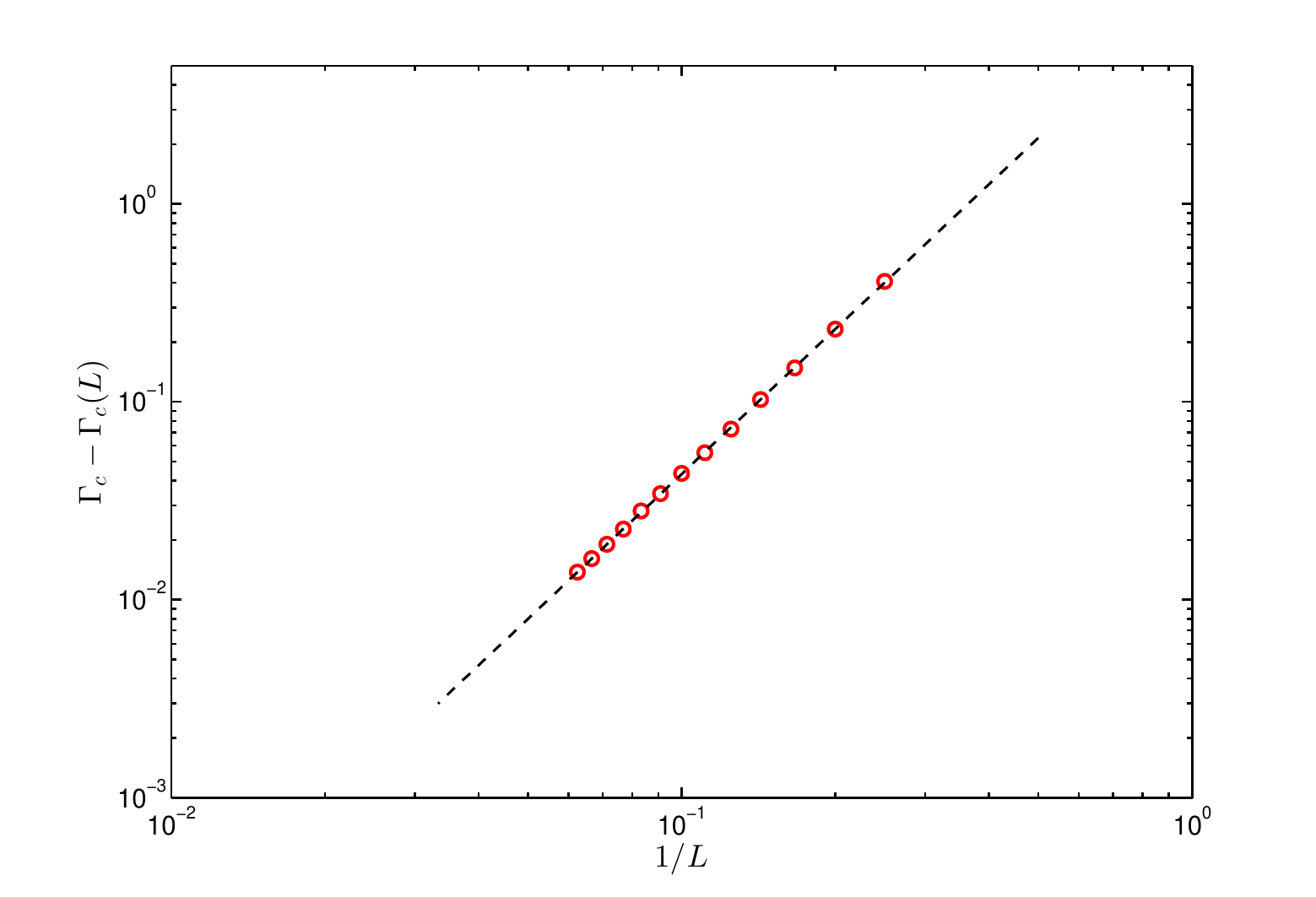}
\end{center}
\caption{ The extrapolation of critical points  $\Gamma_{c}(L)$   of order parameter $M^{x}$ for system size up to $L=16$. We extract the true critical point $\Gamma_{c} \sim 13.845$.}
\label{fig_nuLmax}
\end{figure}

In Fig.\ref{fig_nuLmax}, we show the extrapolation of the critical points $\Gamma_{c}(L)$ and the system size $L$. According to numerical results, the critical points of  thermodynamic limit $L \rightarrow \infty$ is $\Gamma_{c} \sim 13.845$.

\begin{figure}
\begin{center}
\includegraphics[scale=0.35]{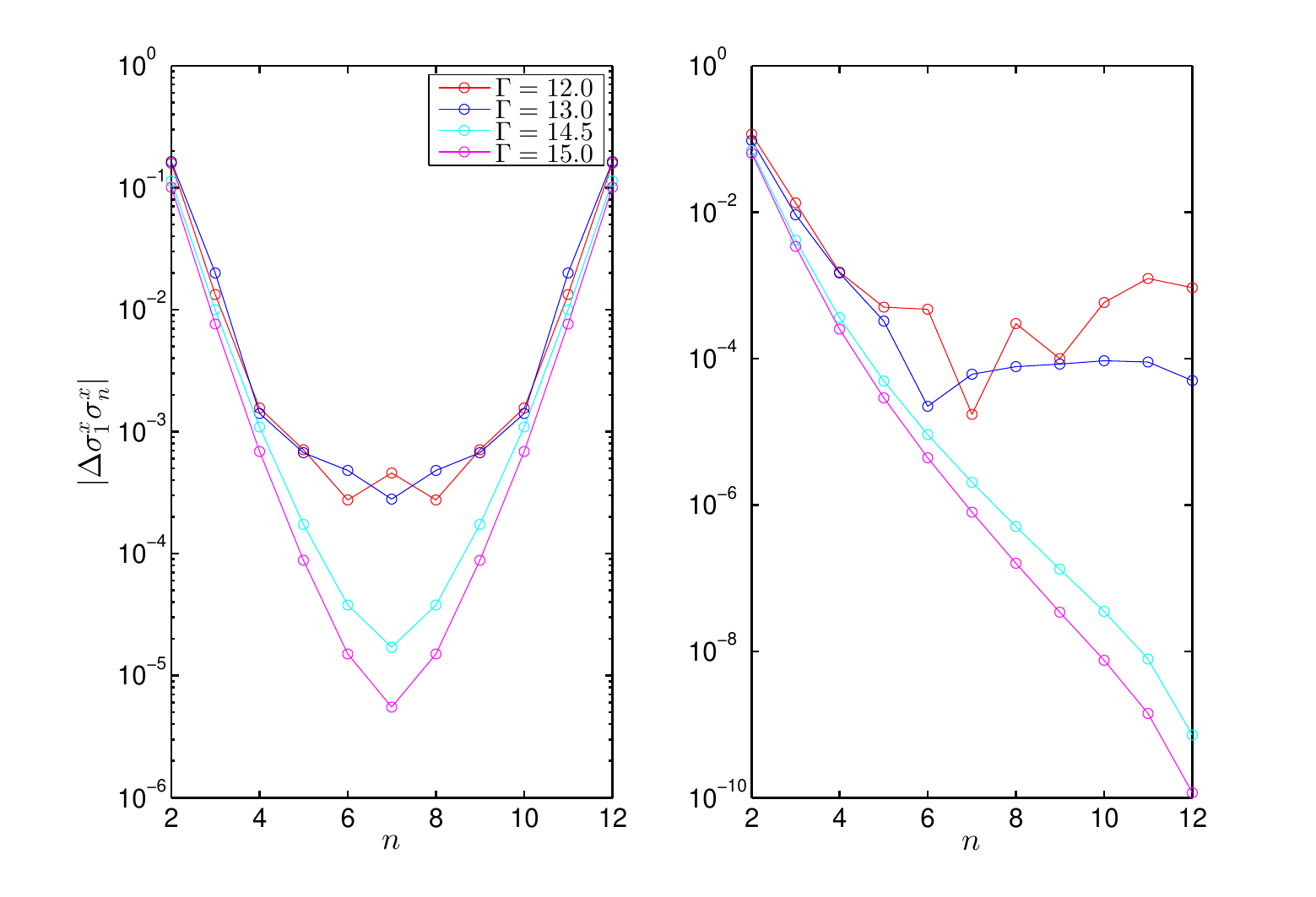}
\end{center}
\caption{Correlation function  for periodic(left) and open(right) boundary. $L = 12$. The correlation length for $\Gamma=14.5$,$\Gamma=15.0$ are $0.5879$,$0.5304$.}
\label{fig_corr_pe&op}
\end{figure}

{\it Correlation. } As seeing Fig.\ref{fig_corr_pe&op}, the correlation function vary with position $n$ for periodic(left panel) and open(right panel) boundary condition.

\section{non-Hermitian case of $L=2$}
We look at the particular and simpler case of $L=2$ spins, which can provide theoretical information about phase transition. It is easy to explicitly write matrix of the effective Hamiltonian
\begin{equation}
H_{\rm eff} = \left(
\begin{array}{cccc}
-i\Gamma &\Omega & \Omega& 0 \\
\Omega & -i\Gamma/2 &0 & 0  \\
\Omega &0& -i\Gamma/2 &0 \\
 0 & 0& 0& 0 \\
\end{array}
\right).
\end{equation}
The eigenvalues are given by
\begin{align*}
&E_{1}=0 \quad (\rm trivial),\\
&E_{2}=-i\Gamma/2, \\
&E_{3,4}=\frac{-i3\Gamma}{4} \pm \frac{1}{4} \sqrt{ 32\Omega^2-\Gamma^2}. 
\end{align*}
Here, we only give out right eigenstates $|\psi_R^i\rangle $ since we use right eigenstates to compute the observables
\begin{eqnarray}
|\psi_R^1 \rangle &=& \left(0,0,0,1 \right)^{T}, \nonumber \\
|\psi_R^2\rangle &=& \left(0,1,-1,0 \right)^{T},  \\
|\psi_R^{3,4}\rangle &=&\frac{1}{\sqrt{2\Omega^2+\Gamma^{2}+|\lambda|^2+2\Gamma Im(\lambda)}}  \\
&  & \left( \sqrt{2}\Omega   ,(\lambda +i\Gamma)/\sqrt{2},(\lambda+i\Gamma)/\sqrt{2},0 \right)^T. \nonumber
\end{eqnarray}
where, for last two wave functions, we remark that eigenvalue $\lambda $ take $E_{3,4}$. The exceptional point(EP) is located at
\begin{equation}
\Gamma_{c}=\sqrt{32} \Omega.
\end{equation}
When the dissipation $\Gamma$ exceed the $\Gamma_{c}$, all non-trivial eigenvalues $E$ become the purely imaginary. Meanwhile, the ground state is $\vert\psi_R^{4}\rangle$ since $E_{4}$ has minimum real part.

The \textit{order parameter}:  $M^{x}=\langle \psi_{R}^{4}\vert   \sigma^{x}\vert  \psi_{R}^{4}\rangle$, where $\sigma^{x}=\sum_{k}\sigma^{k}_{x}$ can be written as matrix form
\begin{equation}
\sigma^{x} = \left(
\begin{array}{cccc}
0 &1 & 1& 0 \\
1 & 0 &0 & 1  \\
1 &0& 0 &1 \\
 0 & 1& 1& 0 \\
\end{array}
\right).
\end{equation}
We obtain the results of $M^{x}$
\begin{equation}
M^{x}=\frac{-\Omega\sqrt{ 32\Omega^2-\Gamma^2}}{2\Omega^2+\Gamma^{2}+|\lambda|^2+2\Gamma Im(\lambda)}.
\end{equation}
When dissipation tends to exceptional point $|\Gamma-\Gamma_{c}| \rightarrow 0$, the order parameter can be written as $M^{x} \sim \sqrt{2 \Gamma_{c}} \sqrt{\Gamma_{c}-\Gamma} \sim \sqrt{\Gamma_{c}-\Gamma}$. Such that the critical exponent $\beta=1/2$

\section{Influence of perturbation to critical exponent}
\begin{figure}
\begin{center}
\includegraphics[scale=0.3]{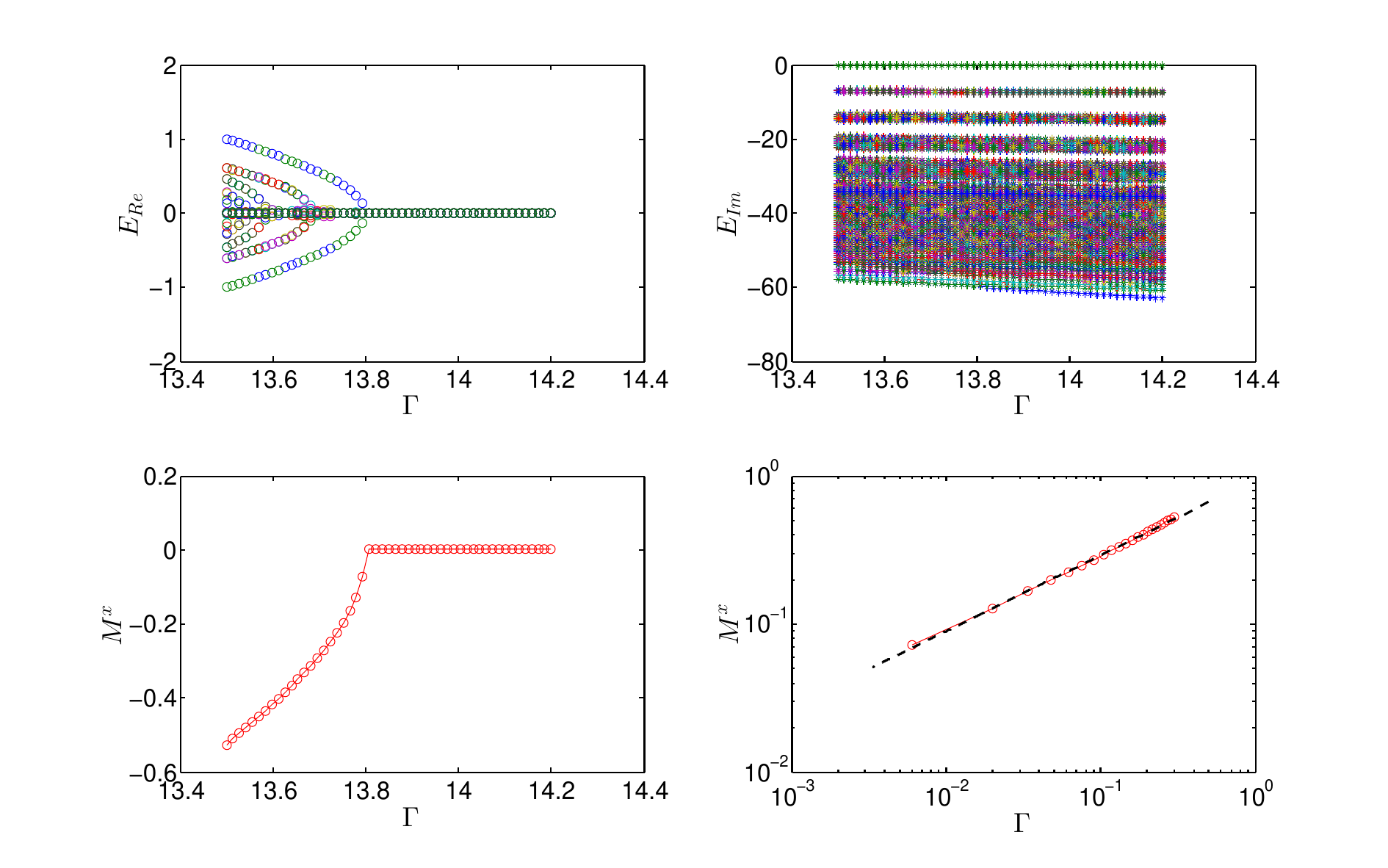}
\end{center}
\caption{The results of one time sampling. The energy(upper subplots) and the order parameter $M^{x}$ of $w=0.1$ near critical point vary with dissipation(lower Left) and its logarithmic scale(lower Right). The critical exponent is extracted as $\beta \sim 0.5118 $. $L=10,\Omega=1$.}
\label{qcp_nc10_Emx_rand_im_w0d1_samp}
\end{figure}

\begin{figure}
\begin{center}
\includegraphics[scale=0.3]{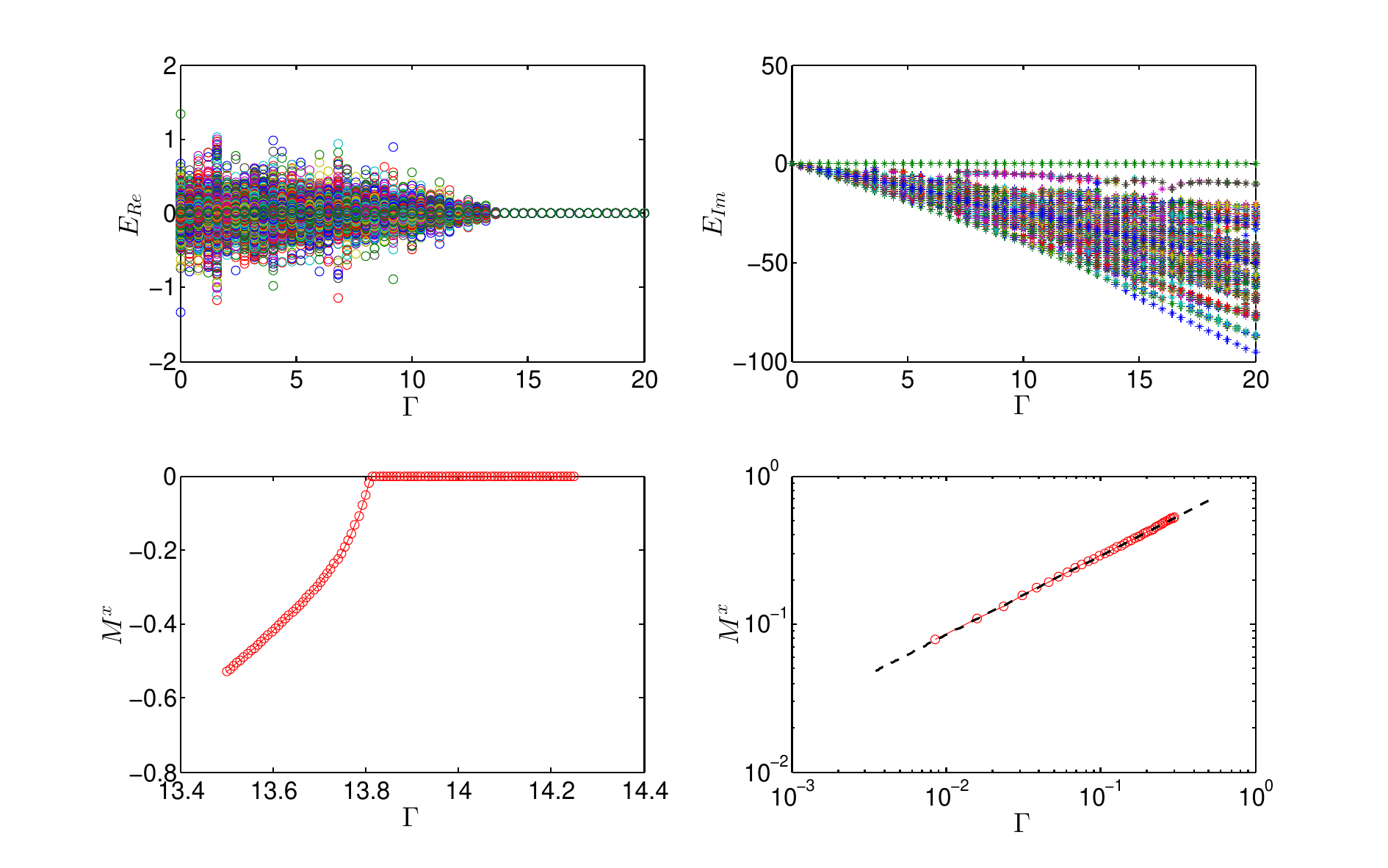}
\end{center}
\caption{Non-Hermitian qcp model of Eq.\ref{Hpert}: The averaged  energy(upper subplots) and the order parameter $M^{x}$ of $w=0.1$ vary with dissipation(lower Left) and its logarithmic scale(lower Right). The critical exponent is extracted as $\beta \sim 0.5348\pm0.0399 $. $L=10,\Omega=1,Nrand=500$.}
\label{qcp_nc10_Emx_rand_im_w0d1_gamm}
\end{figure}

Here, we added imaginary random potential perturbation $z$-direction with Hamiltonian,which can keep energy change from complex to imaginary
 \begin{equation}
\hat H = \Omega \sum_{k}^{L} \left(  \hat \sigma_{x}^{k}\hat \sigma_{n}^{k+1}+\hat \sigma_{n}^{k}\hat \sigma_{x}^{k+1} \right)-\frac{i}{2} \Gamma \sum_{k=1}^{L} \hat \sigma_{+}^{k} \hat \sigma_{-}^{k}+ i\sum_{k=1}^{L} h_{k} \hat \sigma_{+}^{k} \hat \sigma_{-}^{k},
\label{Hpert}
\end{equation}
where random disorder $h_{k} \in [-w/2,w/2]$ and sampling number $Nrand$. We computed the energy  and order parameter $M^{x}$ of ground state of of each sampling as in fig. \ref{qcp_nc10_Emx_rand_im_w0d1_samp}, finally averaged the results of all sampling in fig.\ref{qcp_nc10_Emx_rand_im_w0d1_gamm} as 
\begin{equation}
M^{x}=\frac{1}{Nrand}\sum_{n=1}^{Nrand} \langle\psi_{0}(w)_{n} \vert \hat{M}^{x}\vert \psi_{0}(w)_{n} \rangle.
\end{equation}
 The results of one time sampling in fig.\ref{qcp_nc10_Emx_rand_im_w0d1_samp}, there is  exceptional point and critical exponent is close to $0.5$. After averaging the results over all sampling, order parameter also change from finite value to zeros. Exceptional point is easy to be destroyed by random potential perturbation, see the upper subplots of fig.\ref{qcp_nc10_Emx_rand_im_w0d1_gamm}.  The critical exponent slightly deviates from $0.5$ of fig.\ref{qcp_nc10_Emx_rand_im_w0d1_gamm}(lower subplots) with uncertainty. 


We also consider the special case a two spins system with imaginary random potential, the perturbed Hamiltonian is given as follows,
 \begin{equation}
\hat H = \Omega \left(  \hat \sigma_{x}^{1}\hat \sigma_{n}^{2}+\hat \sigma_{n}^{1}\hat \sigma_{x}^{2} \right)-\frac{i}{2} \Gamma \sum_{k=1}^{2} \hat \sigma_{+}^{k} \hat \sigma_{-}^{k}+ i(h_{1} \hat \sigma_{+}^{1} \hat \sigma_{-}^{1}+h_{2} \hat \sigma_{+}^{2} \hat \sigma_{-}^{2}).
\end{equation}
For convenience, we set $\Omega=1$ as the rescaling unit of system's parameter. The matrix of $H$ is given by 
\begin{equation}
H = \left(
\begin{array}{cccc}
i(\gamma_{1}+\gamma_{2}) &1 & 1& 0 \\
1 & i \gamma_{1}&0 & 0  \\
1 &0& i\gamma_{2} &0 \\
 0 & 0& 0& 0 \\
\end{array}
\right),
\end{equation}
here define $\gamma_{1}=h_{1}-\Gamma/2$ and  $\gamma_{2}=h_{2}-\Gamma/2$. Apart from trivial eigenvalue $E=0$, the characteristic equation of remained three eigenvalues as $|E-H|=0$ satisfy cubic equation
\begin{eqnarray}
&(E-i(\gamma_{1}+\gamma_{2}))(E-i\gamma_{1})(E-i\gamma_{2})-(2E-i(\gamma_{1}+\gamma_{2}))=0& \nonumber \\
&E^3-2i(\gamma_{1}+\gamma_{2})E-[(\gamma_{1}+\gamma_{2})^2+\gamma_{1}\gamma_{2}+2]E& \nonumber \\
&+i[(\gamma_{1}+\gamma_{2})+\gamma_{1}\gamma_{2}(\gamma_{1}+\gamma_{2})]=0&.
\label{cubic}
\end{eqnarray}
Only for balanced perturbation $h_{1}=h_{2}$, above Eq.(\ref{cubic}) can reduce to quadratic equation. Usually random potentials are different that break translation symmetry, eigen-equation changes from square to cubic. The solution formula of cubic equation is too complicated to analysis. The order of characteristic equation is sensitive to the perturbation field. 

\begin{figure}[htbp]
\subfigure{\includegraphics[scale=0.35]{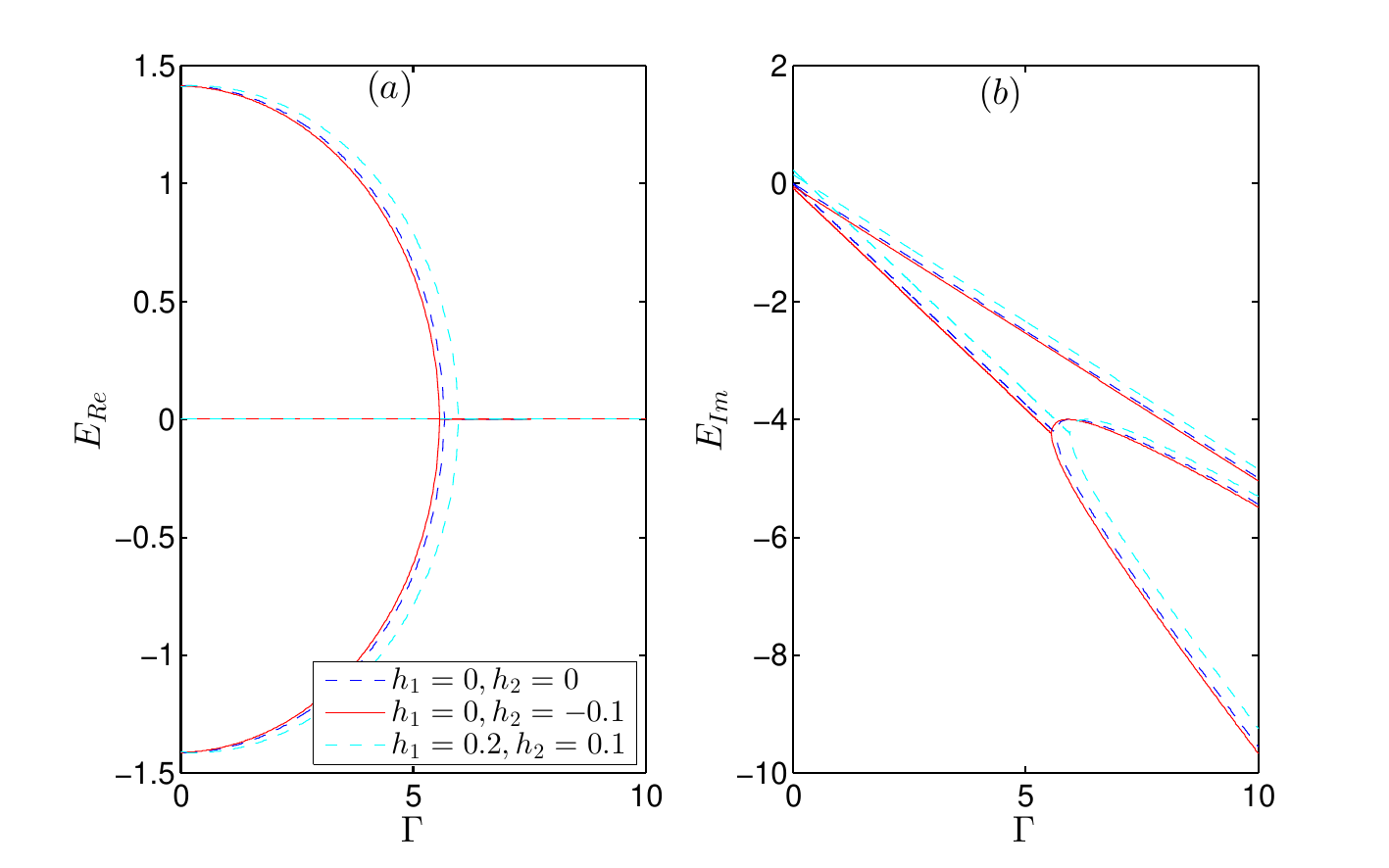}}
\subfigure{\includegraphics[scale=0.35]{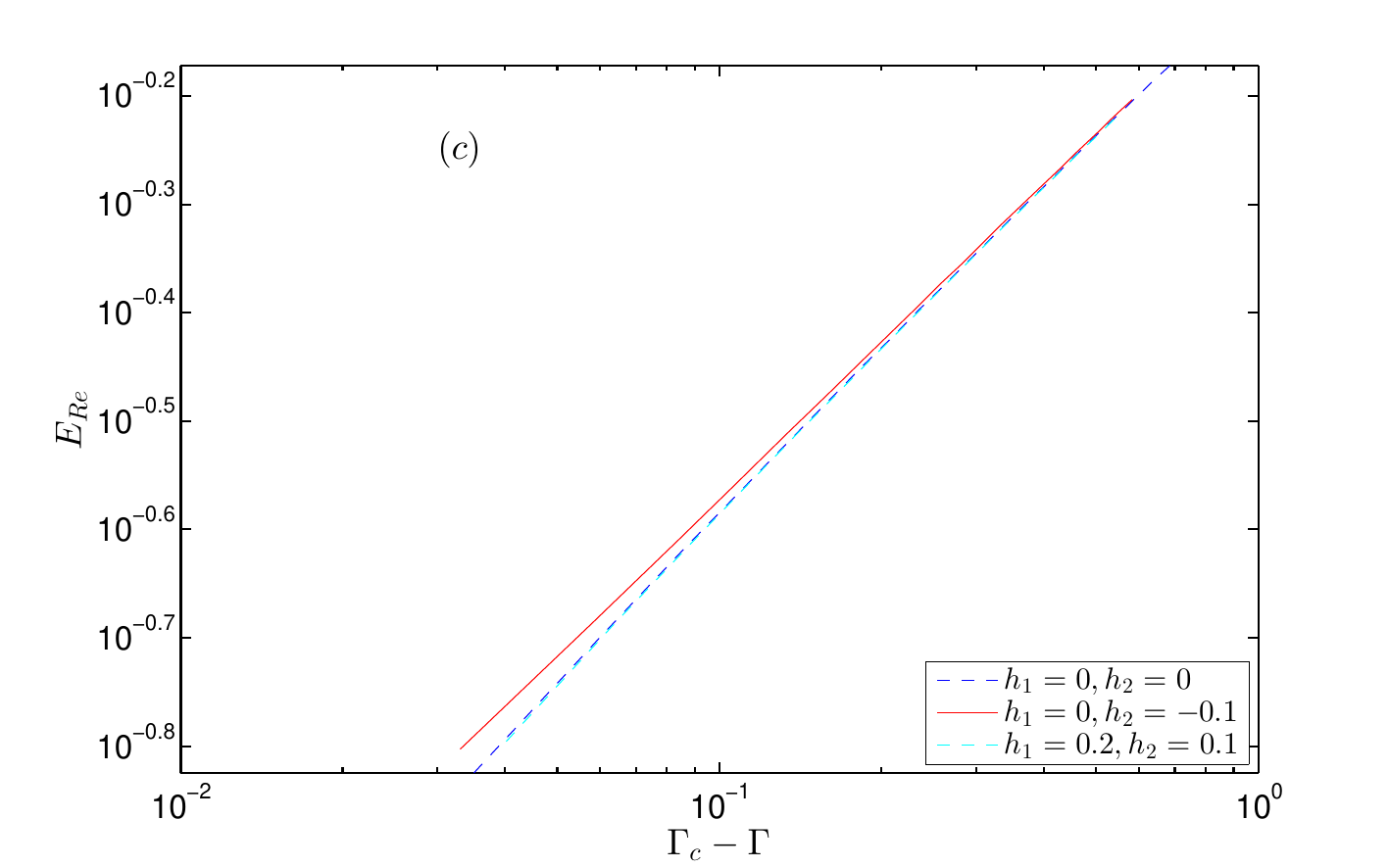}}
\caption{The real(a) and imaginary(b) part of energy.  Blue lines: $h_{1}=h_{2}=0$,  red lines $h_{1}=0,h_{2}=-0.1$ and cyan lines $h_{1}=0.2,h_{2}=0.1$. $L=2,\Omega=1$. (c).  The critical exponents are respectively $\beta \sim [0.5004,4.817,0.5047]$. $L=2,\Omega=1$.}
\label{nc2_qcp_nh_h0d1}
\end{figure}

\begin{figure}
\begin{center}
\includegraphics[scale=0.38]{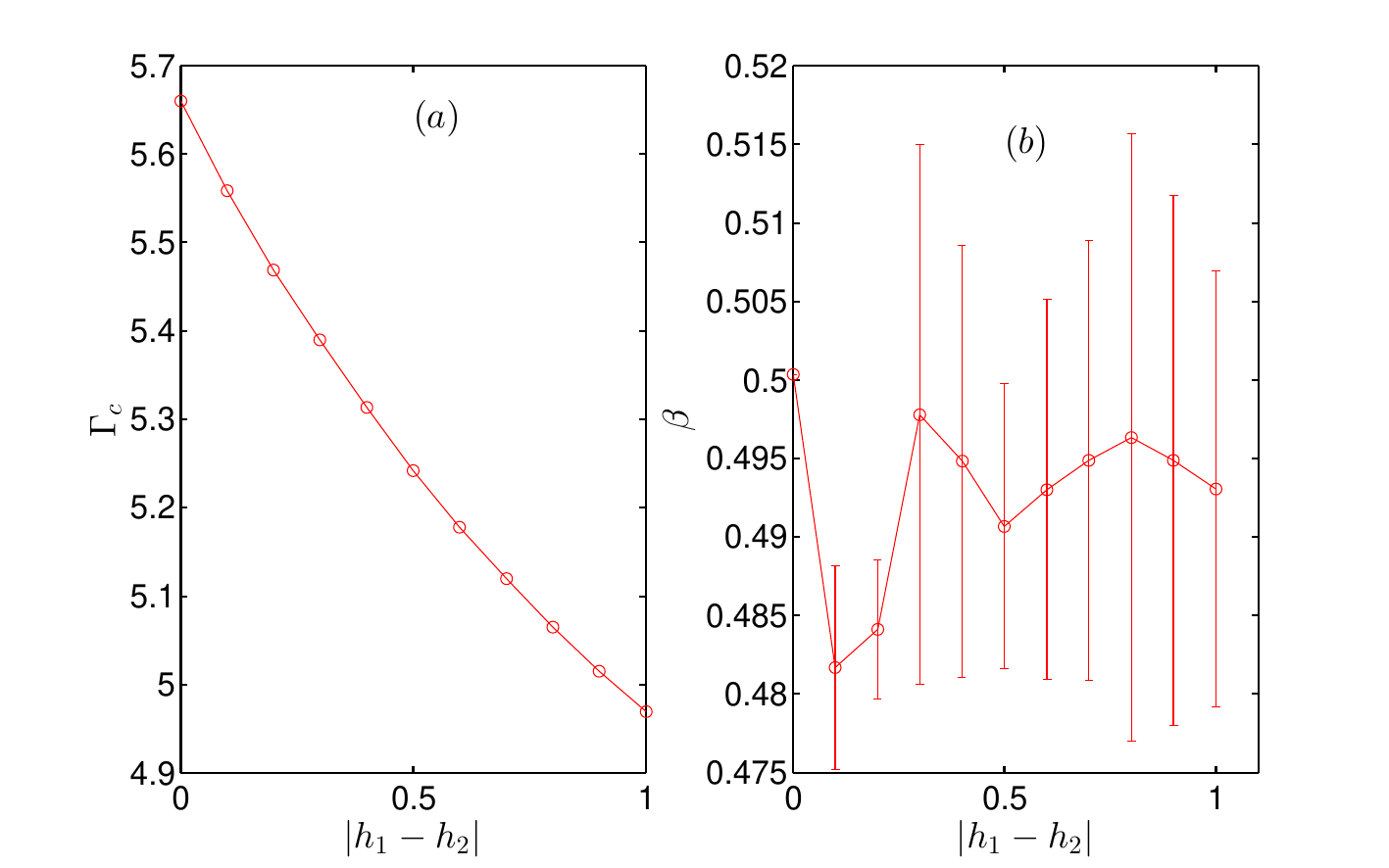}
\end{center}
\caption{The scaling fitting. The critical point(a) and critical exponents $\beta$ (b) vary with field difference.  The parameters $h_{1}=0,h_{2} \in -[0:0.1:1],L=2,\Omega=1$. The errorbar indicate the error of numerical fitting.}
\label{nc2_qcp_nh_h0to1_beta}
\end{figure}

Fig.\ref{nc2_qcp_nh_h0d1}(a)(b) show the real and imaginary parts of the energy vary with dissipation for three sets of perturbation fields. There is tiny difference with the energy between the case without(blue dash line) and with(red solid line and cyan dash line) perturbation. Fig.\ref{nc2_qcp_nh_h0d1}(c) shows the scaling fitting of energy coalescence. The critical exponents of two sets of perturbation  slightly deviate from the results without perturbation. In order understand how perturbation influence the critical exponent, we computed the relation of critical exponent and perturbation field. Fig.\ref{nc2_qcp_nh_h0to1_beta} show the critical point(a) and critical exponent(b) of energy coalescence vary with perturbation field difference $|h_{1}-h_{2}|$. The critical exponents saturate to $0.49$.

\section{Hermitian cases of $L=2$}
It is  easy to explicitly write the Hamiltonian of Eq.(\ref{qcpz}) for simple case $L=2$
\begin{equation}
H= \left(
\begin{array}{cccc}
-\Gamma &\Omega & \Omega& 0 \\
\Omega & -\Gamma/2 &0 & 0  \\
\Omega &0& -\Gamma/2 &0 \\
 0 & 0& 0& 0 \\
\end{array}
\right).
\end{equation}
The eigenvalues are given by 
\begin{align*}
&E_{1}=0\quad  (\rm trivial),\\
&E_{2}=-\Gamma/2, \\
&E_{3,4}=\frac{-3\Gamma}{4} \pm \frac{1}{4} \sqrt{32\Omega^2+\Gamma^2}. 
\end{align*}
If there is an exceptional point, $32\Omega^2+\Gamma^2=32\Omega^2+\Gamma_{re}^{2}-\Gamma_{im}^{2}+i2\Gamma_{re}\Gamma_{im}=0$. We can solve it
\begin{equation}
32\Omega^2+\Gamma_{re}^{2}-\Gamma_{im}^{2}=0, \quad \& \quad    \Gamma_{re}\Gamma_{im}=0.
\end{equation}
Only for $\Gamma_{re}=0$ there is a reasonable solution, but recovers the results of Eq.(9). Through the above analysis, for general non-zero $\Gamma_{\rm re}$ and $\Gamma_{\rm im}$ there are no exceptional points of energy coalescence.

If $\Gamma$ is real value, the eigenvalues
\begin{align*}
&E_{1}=0\quad  (\rm trivial),\\
&E_{2}=-\Gamma_{re}/2, \\
&E_{3,4}=\frac{-3\Gamma_{re}}{4} \pm \frac{1}{4} \sqrt{32\Omega^2+\Gamma_{re}^2}. 
\end{align*}
 eigen-states are given as 
\begin{eqnarray}
|\psi_R^1 \rangle &=& \left(0,0,0,1 \right)^{T}, \nonumber \\
|\psi_R^2\rangle &=& \left(0,1,-1,0 \right)^{T},  \\
|\psi_R^{3,4}\rangle &=&\frac{1}{\sqrt{2\Omega^2+(E+\Gamma_{\rm re})^2}} \nonumber \\
&  & \left(\sqrt{2}\Omega,(E+\Gamma_{\rm re})/\sqrt{2},(E+\Gamma_{\rm re})/\sqrt{2},0 \right)^T. 
\end{eqnarray}
Where $E$ takes $E_{3,4}$ for real $\Gamma$. If $\Gamma$ is real and $\Gamma_{\rm im}=0, \,\Gamma_{\rm re}<0$, let $E_{1}=E_{4}$ for the Hermitian model since $E_{2}>0,\, E_{3}>0$. We obtain that
\begin{equation}
\Gamma_{\rm re}^2=4\Omega^2.
\end{equation}
The solution is $\Gamma_{\rm re}=-2\Omega=\Gamma_{c}$, which is the energy level crossing point for the Hermitian model. Near critical point, $E_{1}<E_{4}$ for $\Gamma_{\rm re}<\Gamma_{c}$, while $E_{4}<E_{1}$ for $\Gamma_{\rm re}>\Gamma_{c}$. 

The order parameter $M^{x}$ of ground state can be given as 
\begin{equation}
M^{x}(\Gamma_{\rm re}<\Gamma_{c})=\langle \psi_{R}^{1}\vert   \sigma^{x}\vert  \psi_{R}^{1}\rangle=0,
\end{equation}
and
\begin{equation}
M^{x}(\Gamma_{\rm re}>\Gamma_{c})=\langle \psi_{R}^{4}\vert   \sigma^{x}\vert  \psi_{R}^{4}\rangle=\frac{2\Omega(E+\Gamma_{\rm re})}{2\Omega^2+(E+\Gamma_{\rm re})^2}.
\end{equation}
which tends to $2\Omega\Gamma_{c}/(2\Omega^2+\Gamma_{c}^2)<0$ for $\Gamma_{\rm re}\rightarrow \Gamma_{c}^{+}$, that the theoretical analysis agree with the results in Fig.\ref{fig_nupMx_hermL}.

\hspace{2em}


\end{document}